\documentclass[11pt]{article}
\usepackage{amssymb,amsfonts,amsmath,amscd}
\usepackage{graphicx}

\makeatletter \@addtoreset{equation}{section} \makeatother

\renewcommand{\theequation}{\thesection.\arabic{equation}}
\renewcommand{\thefootnote}{\alph{footnote}}

\usepackage{theorem}

\newtheorem{theorem}{Theorem}

\newtheorem{thm}{\textbf{Theorem}}
\newtheorem{prop}{\textbf{Proposition}}
\newtheorem{defs}{\textbf{Definition}}[section]
\newtheorem{lem}{\textbf{Lemma}}
\newtheorem{rema}{R e m a r k}

\renewcommand{\theequation}{\thesection.\arabic{equation}}

\numberwithin{equation}{section} 
\numberwithin{theorem}{section}

\newtheorem{aremark}[theorem]{Remark}

\def\dfrac{\displaystyle\frac}
\def\sumd{\displaystyle\sum}
\def\limd{\displaystyle\lim}
\def\intd{\displaystyle\int}
\def\prodd{\displaystyle\prod}

\begin{document}

\title{Bulk Universality for Unitary Matrix Models}

\author{M. Poplavskyi\\Mathematical Division,
B. Verkin Institute for Low Temperature Physics and Engineering\\
National Academy of Sciences of Ukraine\\
47 Lenin Ave., Kharkiv, 61103, Ukraine\\
\smallskip {\rm E-mail:poplavskiymihail@rambler.ru}}

\date{Received April 25, 2008}

\protect\maketitle 
\def\thefootnote{\ifcase\value{footnote}\or {}\or ${\star}$\or ${\star\star}$\or ${\star\!\star\!\star\!}$ \fi}
\newenvironment{proof}{\begin{trivlist}\item[]
{\quad\, P r o o f. }}{\hfill\rule{0.X5em}{0.5em}\end{trivlist}}

\renewcommand{\thesubsection}{\arabic{subsection}.}
\renewcommand{\thethm}{\thesubsection\arabic{thm}}
\renewcommand{\thelem}{\thesubsection\arabic{lem}}
\renewcommand{\theequation}{\thesubsection\arabic{equation}}
\renewcommand{\theprop}{\thesubsection\arabic{prop}}
\renewcommand{\thedefs}{\thesubsection\arabic{defs}}
\renewcommand{\therema}{\thesubsection\arabic{rema}}
\renewcommand{\theexample}{\thesubsection\arabic{example}}
\def\dfrac{\displaystyle\frac}
\def\sumd{\displaystyle\sum}
\def\limd{\displaystyle\lim}
\def\intd{\displaystyle\int}
\def\prodd{\displaystyle\prod}

\begin{abstract}

A proof of universality in the bulk of spectrum of unitary matrix models,
assuming that the potential is globally $C^{2}$ and locally $C^{3}$
function (see Theorem \ref{MainThm}), is given. The proof is based on the
determinant formulas for correlation functions in terms of polynomials
orthogonal on the unit circle. The $sin$-kernel is obtained as a unique
solution of a certain nonlinear integro-differential equation without using
asymptotics of orthogonal polynomials.\vskip3mm

{\em  Key words}: unitary matrix models, local eigenvalue statistics,
uni\-ver\-sality.\smallskip

{\em Mathematics Subject Classification 2000}: 15A52, 15A57.

\end{abstract}\medskip


\setcounter{subsection}{0}

\begin{center}
\subsection{Introduction}
\end{center}


In the paper we study a class of random matrix ensembles known as unitary
matrix models. These models are defined by the probability law
\begin{equation}\label{Model}
p_n\left(U\right)d\mu_n\left(U\right)=Z_{n,2}^{-1} \exp\left\{-n
\hbox{Tr}
V\left(\dfrac{U+U^*}{2}\right)\right\}d\mu_n\left(U\right),
\end{equation}
where $U=\{U_{jk}\}_{j,k=1}^n$ is an $n \times n$ unitary matrix,
$\mu_n\left(U\right)$ is the Haar measure on the group $U(n)$, $Z_{n,2}$ is
the normalization constant and $V:[-1,1] \rightarrow \mathbb{R^+}$ is
a~continuous function called the
potential of the model. %
Denote $e^{i \lambda_j}$ the eigenvalues of unitary matrix $U$.
The joint probability density of $\lambda_j$, corresponding to
\eqref{Model}, is given by (see \cite{Me:91})
\begin{equation}\label{PDef}
p_n \left( \lambda_1, \ldots , \lambda_
n\right)=\dfrac{1}{Z_n}\prodd\limits_{1 \leq j < k \leq n} %
\left| e^{i \lambda_j} -  e^{i \lambda_k}\right|^2 %
\exp \left\{ -n \sum\limits_{j=1}^{n} V \left( \cos\lambda_j
\right) \right\}.
\end{equation}
 To simplify notations, below we will write $V\left(x\right)$
instead of $V\left(\cos x\right)$. Normalized Counting Measure of
eigenvalues (NCM) is given by
\begin{equation}\label{NCM}
N_n\left(\Delta\right)=n^{-1}\sharp \left\{\lambda_l^{(n)} \in
\Delta, \, l=1,\ldots,n\right\}, \quad \Delta \subset [-\pi,\pi].
\end{equation}

The random matrix theory deals with several asymptotic regimes of the
eigenvalue distribution. The global regime is centered around weak
convergence of NCM \eqref{NCM}. Global regime for unitary matrix models was
studied in \cite{Kol:97}. We will use the main result of
\cite{Kol:97}:\medskip

\begin{thm}\label{WC}
Assume that the potential $V$  of the model \eqref{Model} is a
$C^2\left(-\pi,\pi\right)$ function. Then:
\begin{itemize}
\item there exists a measure $N \in \mathcal{M}_1
\left([-\pi,\pi]\right)$ with a compact support $\sigma$ such that
NCM
$N_n$ converges in probability to $N$;%
\item $N$ has a bounded density $\rho$;%
\item denote $\rho_n:=p_1^{(n)}$ the first marginal density, then
for any $\phi \in H^1 \left(-\pi,\pi\right)$
\begin{equation}\label{WConv}
\left|%
\intd\phi\left(\lambda\right)\rho_n\left(\lambda\right)%
\,d \lambda%
-%
\intd\phi\left(\lambda\right)\rho
\left(\lambda\right)%
\,d \lambda%
\right|%
\leq%
C \left\| \phi \right\|^{1/2}_{2} \left\| \phi' \right\|^{1/2}_{2}
n^{-1/2} \ln^{1/2} n,
\end{equation}
where $\left\|\cdot\right\|_2$ denotes $L_2$ norm on $[-\pi,\pi]$
\end{itemize}
\end{thm}\smallskip

One of the main topics of local regime is a universality of local
eigenvalue statistics. Let
\begin{equation}\label{Marginal}
p_l^{\left( n \right)} \left( \lambda_1, \ldots, \lambda_l \right)
= \intd p_n \left( \lambda_1, \ldots, \lambda_l , \lambda_{l+1}
,\ldots ,\lambda_n\right) \, d\lambda_{l+1} \ldots d\lambda_n
\end{equation}
be the $l$-th marginal density of $p_n$.\medskip

\begin{defs}
We call by the bulk of the spectrum the set
\begin{equation}\label{BulkDef}
\left\{ \lambda \in \sigma \, :\, \rho\left(\lambda\right)>0
\right\},
\end{equation}
where $\rho$ is defined in Theorem \ref{WC}.
\end{defs}\smallskip

 The main result of the paper is the proof of
universality conjecture in the bulk of spectrum
\begin{equation}\label{UC}
\lim\limits_{n \to \infty} \left[n
\rho_n\left(\lambda\right)\right]^{-l}%
p_l^{\left( n \right)} \left( \lambda+\dfrac{x_1}{n
\rho_n\left(\lambda\right)}, \ldots, \lambda+\dfrac{x_l}{n
\rho_n\left(\lambda\right)} \right)%
=%
\det \left\{S\left(x_j-x_k\right)\right\}_{j,k=1}^{l},
\end{equation}
where
\begin{equation}\label{sin}
S\left(x\right)=\dfrac{\sin \pi x}{\pi x}.
\end{equation}
By \eqref{UC}, the limiting local distributions of eigenvalues do not
depend on potential $V$ in \eqref{Model}, modulo some weak condition (see
Theorem \ref{MainThm}). The conjecture of universality of all correlation
functions was suggested by F.J.~Dyson (see \cite{Dy:62}) in the early 60s
who proved \eqref{UC}--\eqref{sin} for $V\left(x\right)=0$. First rigorous
proofs for Hermitian matrix models with nonquadratic $V$ appeared only in
the 90s. The case of general $V$ which is locally $C^3$ function was
studied in \cite{Pa-Sh:97}. The case of real analytic potential $V$ was
studied in \cite{De:99}, where the asymptotics of orthogonal polynomials
were obtained. For unitary matrix models the bulk universality was proved
for $V=0$ (see \cite{Dy:62}) and in the case of a linear $V$ (see
\cite{Jo:98}).

%
To prove the main result we need some properties of the polynomials
orthogonal with respect to varying weight on the unit circle. Consider a
system of functions $\left\{e^{i k \lambda}\right\}_{k=0}^{\infty}$ and use
for them the Gram--Schmidt procedure in $L_2\left( \left[-\pi,\pi\right],
e^{-nV\left(\lambda\right)} \right)$. For any $n$ we get the system of
functions
$\left\{P_k^{\left(n\right)}\left(\lambda\right)\right\}_{k=0}^{\infty}$
which are orthogonal and normalized in  $L_2\left( \left[-\pi,\pi\right],
e^{-nV\left(\lambda\right)} \right)$. Since $V$ is even, it is easy to see
that all coefficients of these functions are real. Denote
\begin{equation}\label{Psi}
\psi_k^{\left(n\right)}\left(\lambda\right)=
P_k^{\left(n\right)}\left(\lambda\right)
e^{-nV\left(\lambda\right)/2}.
\end{equation}
Then we obtain the orthogonal in $L_2 (-\pi,\pi)$ functions
\begin{equation}\label{Norm}
\intd\limits_{-\pi}^{\pi} %
\psi_k^{\left(n\right)}\left(\lambda\right) %
\overline{\psi_l^{\left(n\right)}\left(\lambda\right)} %
 \, d \lambda=\delta_{kl}.
\end{equation}
The reproducing kernel of the system \eqref{Psi} is given by
\begin{equation}\label{Kdef}
K_n \left( \lambda, \mu \right) = \sumd\limits_{j=0}^{n-1}
\psi_l^{ \left( n \right) } \left( \lambda \right) %
\overline{\psi_l^{ \left( n \right) } \left( \mu \right)}.
\end{equation}
From \eqref{Norm} we obtain that the reproducing kernel satisfies
the relation
\begin{equation}\label{KOrt}
\intd\limits_{-\pi}^{\pi}
K_n\left(\lambda,\nu\right)K_n\left(\nu,\mu\right) \, d \nu=
K_n\left(\lambda,\mu\right),
\end{equation}
and from the Cauchy inequality we have
\begin{equation}\label{KCB}
\left|K_n\left(\lambda,\mu\right)\right|^2%
\leq %
K_n\left(\lambda,\lambda\right)%
K_n\left(\mu,\mu\right).
\end{equation}
We  also use below the determinant form of the marginal densities
\eqref{Marginal} (see \cite{Me:91})
\begin{equation}\label{KernelProperties}
p_l^{\left( n \right)} \left( \lambda_1, \ldots, \lambda_l \right)
= \dfrac{ \left( n-l \right) ! }{ n!} \hbox{det} \left\| K_n
\left( \lambda_j, \lambda_k \right) \right\|_{j,k=1}^{l}.
\end{equation}
In particular,
\begin{equation}\label{Marg1}
\rho_n\left(\lambda\right)=n^{-1} K_n\left(\lambda,\lambda\right),
\end{equation}
\begin{equation}\label{Marg2}
p_2^{(n)}\left(\lambda,\mu\right)=
\dfrac{K_n\left(\lambda,\lambda\right)K_n\left(\mu,\mu\right)
-\left|K_n\left(\lambda,\mu\right)\right|^2}{n(n-1)}.
\end{equation}
The main result of the paper is\medskip

\begin{thm}\label{MainThm}
Assume that $V\left(\lambda\right)$ is  a $C^2
\left(-\pi,\pi\right)$ function, and there exists an interval
\newline
$\left(a,b\right)~\subset~\sigma$ such that
\begin{equation}\label{cond1}
\sup_{\lambda \in \left(a,b\right)} |V'''\left(\lambda\right)|
\leq C_1 ,\,  \rho\left(\lambda\right) \geq C_2  , \, \lambda \in
\left(a,b\right).
\end{equation}
Then for any $d>0$ and $\lambda_0 \in [a+d,b-d]$ for $K_n$ defined
in \eqref{Kdef} we have
\begin{multline}\label{LimK}
\lim\limits_{n \to \infty}%
\left[K_n\left(\lambda_0,\lambda_0\right)\right]^{-1}%
K_n\left(%
\lambda_0+\dfrac{x}{K_n\left(\lambda_0,\lambda_0\right)},%
\lambda_0+\dfrac{y}{K_n\left(\lambda_0,\lambda_0\right)}%
 \right)%
  \\
 =%
e^{i \left(x-y\right)/2 \rho \left( \lambda_0 \right) }
\dfrac{\sin \pi \left(x-y\right)}{\pi \left(x-y\right)}
\end{multline}
uniformly in $\left(x,y\right)$, varying on a compact set of
$\mathbb{R}^2$.
\end{thm}
\setcounter{rema}{2}

\begin{rema}
{\rm It is easy to see that the universality conjecture \eqref{UC} follows
from Theorem~\ref{MainThm} by \eqref{KernelProperties}.}
\end{rema}

The method of the proof is a version of the one used in \cite{Pa-Sh:97}. An
important part of the proof is a uniform convergence of $\rho_n$ to $\rho$
in a neighborhood of $\lambda_0$:\medskip\setcounter{thm}{3}

\begin{thm}\label{DensityLemma}
Under the assumptions of Theorem~\ref{MainThm} for any $d>0$ there
exists $C\left(d\right)>0$ such that for any $\lambda \in \left[
a+d, b-d \right]$
\begin{equation}\label{StrConv}
\left| \rho_n \left( \lambda \right)-\rho \left( \lambda \right)
\right| \leq C\left( d \right) n^{-2/9}.
\end{equation}
\end{thm}
\setcounter{equation}{0}\setcounter{defs}{0}

\begin{center}
\subsection{Proof of Basic Results}
\end{center}

{P r o o f \ of Theorem~\ref{DensityLemma}.} We will use some facts from
the integral trans\-for\-ma\-tions theory (see \cite{Mu:53}).\medskip

\begin{defs}
Assume that $g\left(\lambda\right)$ is a continuous function on
the interval $\left[-\pi,\pi\right]$. Then its Germglotz
transformation is given by
\begin{equation}\label{GerDef}
F\left[g\right]\left(z\right)=\intd\limits_{-\pi}^{\pi}%
\dfrac{e^{i\lambda}+e^{iz}}{e^{i\lambda}-e^{iz}}
g\left(\lambda\right)\,d\lambda,
\end{equation}
where $z \in \mathbb{C} \backslash \mathbb{R}$.
\end{defs}\smallskip

 The inverse transformation is given by
\begin{equation}\label{GerInv}
g\left(\mu\right)=\dfrac{1}{2\pi}\lim_{z\rightarrow \mu+i0}\Re
F\left[g\right]\left(z\right).
\end{equation}

 For $z = \mu + i \eta$, $\eta \ne 0$, we set
\begin{equation}\label{Fdef}
f_n \left( z \right) = \int\limits_{-\pi}^{\pi} %
\dfrac{ e^{i \lambda} + e^{i z} }{ e^{i \lambda} - e^{i z} } %
\, \rho_n \left( \lambda \right) \, d \lambda.
\end{equation}
Bellow we will derive a "square" equation for $f_n$. Denote
\begin{equation}\label{IDef}
\mathcal{I}_n \left( z \right) = \int\limits_{-\pi}^{\pi} V'(\lambda) %
\dfrac{ e^{i \lambda} + e^{i z} }{ e^{i \lambda} - e^{i z} } %
\, \rho_n \left( \lambda \right) \, d \lambda.
\end{equation}
Integrating by parts in \eqref{IDef},
 from~\eqref{Marginal} we obtain %
\begin{equation*}
\begin{array}{c}
\mathcal{I}_n \left( z \right) = %
\dfrac{1}{Z_n}\int %
V' \left( \lambda_1 \right)%
\!\dfrac{e^{i \lambda_1}+e^{i z}}{e^{i \lambda_1}-e^{i z}} %
\prod\limits_{j<k}\left| e^{i \lambda_j} -  e^{i \lambda_k}\right|^2 %
\!\exp \left\{ -n \sum\limits_{j=1}^{n} V \left( \lambda_j \right)
\right\} \, %
\!\!\prod\limits_{j=1}^{n} d \lambda_j 
\\
=\dfrac{1}{n Z_n}\!\!\int e^{-n V \left( \lambda_1 \right)}
\dfrac{d}{d \lambda_1} %
\!\!\left( %
\!\!\dfrac{e^{i \lambda_1}+e^{i z}}{e^{i \lambda_1}-e^{i z}} %
\prod\limits_{j<k}\left| e^{i \lambda_j} -  e^{i \lambda_k}\right|^2 %
\!\exp\!\left\{ -n\! \sum\limits_{j=2}^{n} V \left( \lambda_j \right)
\right\} \, %
\!\!
\right) %
\!\!\prod\limits_{j=1}^{n} d \lambda_j. 
\end{array}
\end{equation*}
The integrated term equals 0, because all functions here are
$2\pi$~-periodic. After differentiation we have the sum of $n$
terms under integral sign. Denote
\begin{gather*}
I_0 \left( z \right)= %
\dfrac{1}{n Z_n}\int
\dfrac{d}{d \lambda_1} %
\left( %
\dfrac{e^{i \lambda_1}+e^{i z}}{e^{i \lambda_1}-e^{i z}} %
\right)%
\prod\limits_{j<k}\left| e^{i \lambda_j} -  e^{i \lambda_k}\right|^2 %
\exp \left\{ -n \sum\limits_{j=1}^{n} V \left( \lambda_j \right)
\right\} \, %
\prod\limits_{j=1}^{n} d \lambda_j ,%
\end{gather*}
\begin{gather*}
I_m \left( z \right)= %
\dfrac{1}{n Z_n}\int %
\dfrac{e^{i \lambda_1}+e^{i z}}{e^{i \lambda_1}-e^{i z}} %
\prod\limits_{2 \leq j < k \leq n} %
\left| e^{i \lambda_j} -  e^{i \lambda_k}\right|^2 %
\dfrac{d}{d \lambda_1} \left| e^{i \lambda_1} -  e^{i
\lambda_m}\right|^2 %
\\
\times %
\prod\limits_{k \ne m} %
\left| e^{i \lambda_1} -  e^{i \lambda_k}\right|^2 %
\exp \left\{ -n \sum\limits_{j=1}^{n} V \left( \lambda_j\right)\right\} \, %
\prod\limits_{j=1}^{n} d \lambda_j, \quad%
m=\overline{2,n}.
\end{gather*}
From symmetry with respect to $\lambda_j$ we obtain that all $I_m
\left( z \right)$, except $I_0(z)$, are equal, hence
\begin{equation*}
\mathcal{I}_n \left( z \right)=I_0\left( z \right)+\left( n-1
\right) I_2\left( z \right).
\end{equation*}
\begin{align*}
I_0 \left( z \right)& = %
\dfrac{1}{n} \intd\limits_{-\pi}^{\pi} %
\dfrac{d}{d \lambda_1} %
\left( %
\dfrac{e^{i \lambda_1}+e^{i z}}{e^{i \lambda_1}-e^{i z}} %
\right)%
\rho_n \left( \lambda_1 \right) \, d \lambda_1 %
\\
&= %
-\dfrac{2i}{n} \intd\limits_{-\pi}^{\pi} %
\!\!\dfrac{ e^{i \lambda_1} e^{i z} }{ \left( e^{i \lambda_1} -e^{i z}
\right)^2 }
\rho_n \left( \lambda_1 \right) d \lambda_1 %
=%
-\dfrac{i}{2n} \intd\limits_{-\pi}^{\pi} %
\!\!\left(
\dfrac{e^{i \lambda_1}+e^{i z}}{e^{i \lambda_1}-e^{i z}} %
\right)^2
\!\rho_n \left( \lambda_1 \right) d \lambda_1 %
+\dfrac{i}{2n}.
\end{align*}
To transform $I_2$, we use the symmetry of $p_2^{ \left( n \right
)}$ ( $ p_2^{ \left( n \right )}\left( \lambda_1, \lambda_2
\right) = p_2^{ \left( n \right )} \left( \lambda_2, \lambda_1
\right)$ ).
\begin{align*}
I_2 \left( z \right)& =%
\dfrac{1}{n}\int %
\dfrac{e^{i \lambda_1}+e^{i z}}{e^{i \lambda_1}-e^{i z}} %
\dfrac{ %
\dfrac{d} %
{d \lambda_1} \left| e^{i \lambda_1} -e^{i\lambda_2}\right|^2 %
} %
{ %
\left| e^{i \lambda_1} -e^{i\lambda_2}\right|^2
}%
\, p_2^{ \left ( n \right ) } \left( \lambda_1,\lambda_2\right) \, %
d \lambda_1 d \lambda_2 %
\\
&= %
\dfrac{i}{n}\int %
\dfrac %
{e^{i \lambda_1}+e^{i z}} %
{e^{i \lambda_1}-e^{i z}} %
\dfrac %
{ e^{i \lambda_1}+e^{i \lambda_2}} %
{ e^{i \lambda_1}-e^{i \lambda_2}} %
\, p_2^{ \left ( n \right ) } \left( \lambda_1,\lambda_2\right) \, %
d \lambda_1 d \lambda_2 %
\\
&= %
\dfrac{i}{2n}\int %
\left( %
\dfrac{e^{i \lambda_1}+e^{i z}}{e^{i \lambda_1}-e^{i z}} -%
\dfrac{e^{i \lambda_2}+e^{i z}}{e^{i \lambda_2}-e^{i z}} %
\right) %
\dfrac %
{ e^{i \lambda_1}+e^{i \lambda_2}} %
{ e^{i \lambda_1}-e^{i \lambda_2}}%
\, p_2^{ \left ( n \right ) } \left( \lambda_1,\lambda_2\right) \, %
d \lambda_1 d \lambda_2 %
\\
&= %
-\dfrac{i}{2n}\int %
\dfrac
{%
2 \left( e^{i \lambda_1}+ e^{i \lambda_2} \right) e^{i z}
}%
{%
\left( e^{i \lambda_1}-e^{i z} \right) %
\left( e^{i \lambda_2}-e^{i z} \right) %
}%
\, p_2^{ \left ( n \right ) } \left( \lambda_1,\lambda_2\right) \, %
d \lambda_1 d \lambda_2 %
\\
&= %
\dfrac{i}{2n}-\dfrac{i}{2n} \int %
\dfrac{e^{i \lambda_1}+e^{i z}}{e^{i \lambda_1}-e^{i z}} %
\dfrac{e^{i \lambda_2}+e^{i z}}{e^{i \lambda_2}-e^{i z}} %
\, p_2^{ \left ( n \right ) } \left( \lambda_1,\lambda_2\right) %
\, d \lambda_1 d \lambda_2. %
\end{align*}
Therefore, from~\eqref{Marginal} and~\eqref{KernelProperties} we
obtain
\begin{equation}\label{Ieq}
\mathcal{I}_n \left( z \right) = \dfrac{i}{2}- %
\dfrac{i}{2} f_n^2 \left( z \right)-%
\dfrac{i}{n^2} \int %
\left| K_n\left( \lambda_1, \lambda_2 \right ) \right|^2 %
\dfrac %
{ \left( e^{i \lambda_1}-e^{i \lambda_2} \right)^2 e^{2 i z} } %
{
\left( e^{i \lambda_1}-e^{i z} \right)^2 %
\left( e^{i \lambda_2}-e^{i z} \right)^2 %
}
\, d \lambda_1 d \lambda_2. %
\end{equation}
On the other hand, denoting
\begin{equation}\label{Qdef}
Q_n \left( z \right)=\intd\limits_{-\pi}^{\pi} %
\dfrac%
{%
e^{i \lambda}+e^{i z}
}%
{%
e^{i \lambda}-e^{i z}
}%
\left(V' \left( \lambda \right) - V' \left ( \mu \right)\right)
 \, \rho_n
\left( \lambda \right) \, d \lambda,
\end{equation}
for $z=\mu+i\eta$, from~\eqref{Fdef} we get
\begin{equation}\label{IQ}
\mathcal{I}_n \left(z\right)=Q_n
\left(z\right)+V'\left(\mu\right)f_n\left(z\right).
\end{equation}
Finally, from~\eqref{Ieq} and~\eqref{IQ} we obtain the "square"
equation
\begin{equation}\label{SquareEq}
f^2_n \left( z \right) - 2iV'\left( \mu \right)f_n \left( z
\right)-2iQ_n \left( z \right) -1=-\dfrac{2}{n^2} G_n\left( z
\right),
\end{equation}
with
\begin{equation*}
G_n \left( z \right)= %
\int %
\left| K_n\left( \lambda_1, \lambda_2 \right ) \right|^2 %
\dfrac %
{ \left( e^{i \lambda_1}-e^{i \lambda_2} \right)^2 e^{2 i z} } %
{
\left( e^{i \lambda_1}-e^{i z} \right)^2 %
\left( e^{i \lambda_2}-e^{i z} \right)^2 %
}
\, d \lambda_1 d \lambda_2 .%
\end{equation*}
To proceed further we have to prove the following properties of the
reproducing kernel $K_n$.\medskip

\begin{lem}\label{LemmaPolinoms}
Let $K_n\left( \lambda , \mu \right)$ be defined by~\eqref{Kdef}.
Then under the conditions of Theorem~\ref{MainThm} for any $\delta
> 0$
\begin{gather}
\left| %
\intd  \left( e^{i \lambda} - e^{i \mu} \right) %
\left| K_n\left( \lambda , \mu \right) \right|^2 %
\, d \mu %
\right|%
\leq \dfrac{1}{2}%
\left[
\left| \psi_{n-1}^{ \left( n \right) } \left( \lambda \right) \right|^2%
+
\left| \psi_n^{ \left( n \right) } \left( \lambda \right) \right|^2%
\right]
,%
\label{1OneVariable}
\\
\intd  %
\left| e^{i \lambda} - e^{i \mu} \right|^2 %
\left| K_n\left( \lambda , \mu \right) \right|^2 %
\, d \mu %
\leq %
\left[
\left| \psi_{n-1}^{ \left( n \right) } \left( \lambda \right) \right|^2%
+
\left| \psi_n^{ \left( n \right) } \left( \lambda \right) \right|^2%
\right]
,%
\label{2OneVariable}
\\
\intd \left| e^{i \lambda } - e^{i \mu }\right|^2 %
\left| K_n\left( \lambda , \mu \right) \right|^2 %
\, d \lambda d \mu \leq 2 %
,%
\label{Variation}
\end{gather}
\begin{gather}
\intd\limits_{\left| e^{i \lambda}-e^{i \mu} \right| > \delta} %
\left| K_n\left( \lambda , \mu \right) \right|^2 %
\, d \mu \leq \delta^{-2} %
\left[
\left| \psi_{n-1}^{ \left( n \right) } \left( \lambda \right) \right|^2%
+
\left| \psi_n^{ \left( n \right) } \left( \lambda \right) \right|^2%
\right]
,%
\label{IntDist1}
\\
\intd\limits_{\left| e^{i \lambda}-e^{i \mu} \right| > \delta} %
\left| K_n\left( \lambda , \mu \right) \right|^2 %
\, d \lambda d \mu \leq 2 \delta^{-2}.%
\label{IntDist2}
\end{gather}
\end{lem}\smallskip

 It is easy to see that $\left|e^{i \lambda}-e^{i
z}\right|> C\left|\eta\right|$ if $\left|\eta\right|<1$ for some
$C>0$. Hence, from~\eqref{Variation} and~\eqref{SquareEq} we
derive
\begin{equation}\label{SquareEq2}
f^2_n \left( z \right) - 2iV'\left( \mu \right)f_n \left( z
\right)-2iQ_n \left( z \right) -1=O\left(n^{-2}\eta^{-4}\right).
\end{equation}\medskip

\begin{lem}\label{RhoEstimation}
Under the conditions of Theorem~\ref{MainThm} for any $d>0$ and
$\lambda \in \left[a+d,b-d\right]$
\begin{gather}
\rho_n \left( \lambda \right) \leq C, %
\label{Rho_bound}
\\
\left| \dfrac{d\rho_n\left( \lambda \right)}{d\lambda} \right| \leq%
C_1 \left(%
\left| \psi_{n}^{ \left( n \right) } \left( \lambda \right) \right|^2%
+%
\left| \psi_{n-1}^{ \left( n \right) } \left( \lambda \right) \right|^2%
\right)+C_2.%
\label{DRho_bound}
\end{gather}
\end{lem}\smallskip

>From the conditions of Theorem~\ref{MainThm}, we obtain that
$V''\left(\lambda\right)$ is bounded on the interval
$\left[a,b\right]$. Hence, for $\mu \in \left[a+d,b-d\right]$ and
sufficiently small $\eta$  we have
\begin{multline}\label{Q_n}
\left|%
Q_n \left( \mu + i \eta \right) - Q_n \left( \mu \right)%
\right|%
\leq%
\left| e^{- \eta} -1 \right|%
\intd\limits_{-\pi}^{\pi} %
\dfrac%
{%
\left| V'\left( \lambda \right) - V'\left( \mu \right) \right|
\rho_n\left( \mu \right)
}%
{%
\left|e^{i \lambda}-e^{i \mu}\right| %
\left|e^{i \lambda}-e^{i z}\right| %
}%
\, d\lambda 
\\
\leq %
C \eta %
\left(\,\,%
\intd\limits_{\left| \lambda - \mu \right|<d/2} %
\dfrac{d \lambda }{\left| \left(
\lambda -  \mu \right)^2+\eta^2\right|^{1/2} }%
+%
\intd\limits_{\left|  \lambda- \mu \right|>d/2} %
\dfrac{\rho_n\left(\lambda\right)\, d\lambda}%
{%
\left| \left( \lambda - \mu \right)^2+\eta^2\right|^{1/2} %
\left|\lambda - \mu \right|
}%
\right)%
\\
\leq%
C \eta \ln^{-1}\eta+C \eta d^{-2} \leq C \eta \ln^{-1}\eta.
\end{multline}
Besides, applying~\eqref{WConv},
for $\phi\left(\lambda\right)=%
\dfrac%
{%
e^{ i \lambda} + e^{i \mu}
}%
{%
e^{ i \lambda} - e^{i \mu}
}%
\left(%
V'(\lambda)-V'(\mu)
\right)$ we get
\begin{equation}\label{Q_nQ}
Q_n\left(\mu\right)=Q\left(\mu\right)+O\left(n^{-1/2} \ln^{1/2} n
\right),
\end{equation}
where
\begin{equation}\label{Qdefine}
Q\left(\mu\right)=\intd\limits_{-\pi}^{\pi}
\dfrac%
{%
e^{i\lambda}+e^{i\mu}
}%
{%
e^{i\lambda}-e^{i\mu}
}%
\left(%
V'\left(\lambda\right)-V'\left(\mu\right)
\right)%
\rho\left(\lambda\right)%
\,d\lambda.
\end{equation}
Combining \eqref{Q_n} and \eqref{Q_nQ}, we find
\begin{equation}\label{Q_nQ2}
Q_n\left(\mu+i\eta\right)=Q\left(\mu\right)+O\left( \eta \ln^{-1}
\eta \right) + O\left( n^{-1/2} \ln^{1/2} n \right).
\end{equation}
From~\eqref{Q_nQ2} and~\eqref{SquareEq2} for $z=\mu+in^{-4/9}$ we have
\begin{equation}\label{sqeqf}
f_n^2\left(z\right)-2iV'\left(\mu\right)f_n\left(z\right)-
2iQ\left(\mu\right)-1=O(n^{-2/9}).
\end{equation}\smallskip

\begin{lem}\label{RhoRep}
\begin{equation}\label{SqRep}
\rho \left( \mu \right)
=\dfrac{1}{2\pi}\sqrt{2iQ\left(\mu\right)+1-
\left(V'\left(\mu\right)\right)^2}.
\end{equation}
\end{lem}\smallskip

Lemma~\ref{RhoRep} and the equation~\eqref{sqeqf} imply that for
$z=\mu+in^{-4/9}$
\begin{equation}\label{fConv}
\dfrac{1}{2\pi}\Re f_n\left(z\right)=\rho\left(\mu\right)+
O\left(n^{-2/9}\right)\rho^{-1}\left(\mu\right).
\end{equation}

\begin{lem}\label{IntRhoEst}
For $d>0$, $k=n-1,n$ and $\mu \in \left[a+d,b-d\right]$
\begin{gather}
\intd\limits_{\left|\lambda-\mu\right|<n^{-1/4}} %
\left| \psi_k^{\left(n\right)}\left(\lambda\right)\right|^2 %
\, d \lambda \leq C n^{-1/4}, %
\label{IntRhoE}
\\
\left| \psi_k^{\left(n\right)}\left(\lambda\right)\right|^2 %
\leq C n^{7/8}, \, \left|\mu-\lambda\right|\leq n^{-1/4}.
\label{PolEst}
\end{gather}
\end{lem}\smallskip

Taking into account~\eqref{fConv}, to prove
Theorem~\ref{DensityLemma} it is enough to show that
$\dfrac{1}{2\pi}\Re
f_n\left(z\right)=\rho_n\left(\mu\right)+O\left(n^{-2/9}\right)$.
We use an evident relation
\begin{equation*}
\Re \dfrac{e^{ i \lambda}+e^{ i z}}{e^{ i \lambda}-e^{ i z}}=
\dfrac{ \sinh \eta}{\cosh \eta - \cos \left(\lambda-\mu\right)}=
\dfrac{d}{d\lambda} 2 \arctan
\left(\tan\left(\dfrac{\lambda-\mu}{2}\right) \coth
\left(\dfrac{\eta}{2}\right)\right).
\end{equation*}
Combining the relation $\dfrac{1}{2\pi} \intd \Re \dfrac{e^{i \lambda} +
e^{i z}}{e^{i \lambda} - e^{i z}}d\lambda =1$ with~\eqref{Rho_bound}, we
obtain
\begin{multline*}
\hskip4,5cm\left|\dfrac{1}{2\pi} f_n\left(z\right)-
\rho_n\left(\mu\right)\right|
\\
=\left(2\pi\right)^{-1}%
\Biggl|%
\Biggl(\,\,
\intd\limits_{\left|\mu-\lambda\right|\leq \eta^{1/2}}%
+
\intd\limits_{\eta^{1/2}\leq\left|\mu-\lambda\right|\leq d/2}%
+
\intd\limits_{\left|\mu-\lambda\right|\geq d/2}%
\Biggr)%
\dfrac{ \sinh \eta}{\cosh \eta - \cos \left(\lambda-\mu\right)}%
\\
\times\left(\rho_n\left(\lambda\right)-\rho_n\left(\mu\right)\right) \, d
\lambda \Biggr|
\\
\leq C%
\Biggl|\, %
\intd\limits_{\left|s\right|\leq \eta^{1/2}}%
\dfrac{ \sinh \eta}{\cosh \eta - \cos s}%
\left(\rho_n\left(s+\mu \right)-\rho_n\left(\mu\right)\right) \,
ds %
\Biggr|%
+%
C\eta^{1/2}%
+%
C\eta.\qquad\quad
\end{multline*}
Using~\eqref{DRho_bound} and~\eqref{IntRhoE}, we get finally
\begin{equation*}
\left|\dfrac{1}{2\pi} f_n\left(z\right)-
\rho_n\left(\mu\right)\right| \leq
C %
\intd\limits_{\left|s\right|<\eta^{1/2}} \left|
\rho_n'\left(\mu+s\right)\right| d s %
+ C \eta^{1/2}
\leq%
C \eta^{1/2}.
\end{equation*}
Theorem~\ref{DensityLemma} is proved. \hfill\rule{0.5em}{0.5em}\smallskip

Now we pass to the proof of Theorem~\ref{MainThm}. We will use the
following representation of $K_n$, which can be derived from the
well-known identities of random matrix theory (see \cite{Me:91})
\begin{multline}\label{KInt}
\dfrac{1}{n}K_n\left(\lambda,\mu\right)=
\dfrac{1}{n}\sumd\limits_{j=0}^{n-1}
\psi_l^{ \left( n \right) } \left( \lambda \right) %
\overline{\psi_l^{ \left( n \right) } \left( \mu \right)}%
=%
Q_{n,2}^{-1} e^{-n\left( V\left(\lambda\right)+V\left(\mu\right)
\right)/2}
\\
\times\intd %
\prodd\limits_{j=2}^{n} \left(e^{i\lambda}-e^{i\lambda_j}\right)
\left(e^{-i\mu}-e^{-i\lambda_j}\right)
e^{-nV\left(\lambda_j\right)} d\lambda_j \prodd\limits_{2 \leq j <
k \leq n} \left|e^{i\lambda_j}-e^{i\lambda_k}\right|^2,
\end{multline}
where $Q_{n,2}=n!\prodd\limits_{j=0}^{n-1}%
\left|\gamma_l^{\left(n\right)}\right|^{-2}$, and $\gamma _{l}^{(n)}$ is
the coefficient in front of $e^{i l \lambda }$ in the function
$P_{l}^{(n)}$.\smallskip\setcounter{rema}{4}\setcounter{lem}{5}

\begin{rema}
{\rm Consider the determinant (see~\eqref{PDef})
\begin{equation*}
\det%
\left\{%
e^{i k \lambda_j}
\right\}_{k,j=0}^{n-1}%
=e^{i (n-1)\sum \lambda_j /2}%
\det%
\left\{%
e^{i (k-(n-1)/2) \lambda_j}
\right\}_{k,j=0}^{n-1}.%
\end{equation*}
Taking the complex conjugate, we obtain
\begin{align*}
\overline{
\det%
\left\{%
e^{i k \lambda_j}
\right\}_{k,j=0}^{n-1}%
}
&=e^{-i (n-1) \sum \lambda_j /2}%
\det%
\left\{%
e^{-i (k-(n-1)/2) \lambda_j}
\right\}_{k,j=0}^{n-1}%
\\
&=(-1)^{[n/2]}
e^{-i (n-1)\sum \lambda_j /2}%
\det%
\left\{%
e^{i (k-(n-1)/2) \lambda_j}
\right\}_{k,j=0}^{n-1}.%
\end{align*}
Thus, from~\eqref{KInt} we get that the function
$e^{-i(n-1)(\lambda-\mu)/2}K_n\left(\lambda,\mu\right)$ is real valued.}
\end{rema}\smallskip

 Now denote
\begin{equation}\label{KerDef}
\mathcal{\widetilde{K}}_n\left(x,y\right)=%
\dfrac{1}{n} K_n
\left(%
\lambda_0+\dfrac{x}{n},%
\lambda_0+\dfrac{y}{n}%
\right),%
\qquad%
\mathcal{K}_n\left(x,y\right)=%
e^{-i(n-1)(x-y)/2n}\mathcal{\widetilde{K}}_n\left(x,y\right).%
\end{equation}
From the above we have that $\mathcal{K}_n(x,y)$ is a real-valued and
symmetric function. We get from \eqref{Kdef}--\eqref{KCB}
\begin{gather}
\intd\limits_{-n\pi}^{n\pi} \mathcal{K}_n\left(x,z\right)%
\mathcal{K}_n\left(z,y\right)%
d z=\mathcal{K}_n\left(x,y\right)%
, \quad %
\left|\mathcal{K}_n\left(x,y\right)\right|^{2} \leq
\mathcal{K}_n\left(x,x\right)
\mathcal{K}_n\left(y,y\right),%
\label{KerOrt}
\\
\mathcal{K}_n\left(x,x\right)=\rho_n\left(\lambda_0+x/n\right) \leq C%
, \quad \left|\mathcal{K}_n\left(x,y\right)\right| \leq C %
, \quad \mbox{for} \left|x\right|,\,\left|y\right|\leq nd_0/2
\label{KerEst}
\end{gather}
Differentiating in~\eqref{KInt}
$\mathcal{\widetilde{K}}_n\left(x,y\right)$ with respect to $x$
for $\lambda=\lambda_0+x/n,\,\mu=\mu_0+y/n$, we get
\begin{gather*}
\dfrac{\partial}{\partial x}\mathcal{\widetilde{K}}_n\left(x,y\right)=%
-\dfrac{1}{2}V'\left(\lambda\right)\mathcal{\widetilde{K}}_n\left(x,y\right)
+ \dfrac{n-1}{Q_{n,2}}e^{-n\left( V\left(\lambda\right)+V\left(\mu\right)
\right)/2}%
\\
\times\intd %
\dfrac{ie^{i\lambda}}{e^{i\lambda}-e^{i\lambda_2}} \prodd\limits_{j=2}^{n}
\left(e^{i\lambda}-e^{i\lambda_j}\right)
\left(e^{-i\mu}-e^{-i\lambda_j}\right) d\lambda_j %
\prodd\limits_{2 \leq j < k \leq n}
\left|e^{i\lambda_j}-e^{i\lambda_k}\right|^2
\\
=%
-\dfrac{1}{2}V'\left(\lambda\right)\mathcal{\widetilde{K}}_n\left(x,y\right)
\\
+\dfrac{i}{n^2}\intd\limits_{-\pi}^{\pi} %
\dfrac{e^{i \lambda}}{e^{i \lambda}-e^{i\lambda_2}}%
\left(%
K_n\left(\lambda_2,\lambda_2\right)K_n\left(\lambda,\mu\right)-%
K_n\left(\lambda,\lambda_2\right)K_n\left(\lambda_2,\mu\right)
\right)%
\,d \lambda_2
\\
=%
-\dfrac{1}{2}V'\left(\lambda\right)\mathcal{\widetilde{K}}_n\left(x,y\right)
\end{gather*}
\begin{multline}\label{KerDiff}
\dfrac{i}{2n^2}\intd\limits_{-\pi}^{\pi} %
\dfrac{e^{i \lambda}+e^{i\lambda_2}}{e^{i \lambda}-e^{i\lambda_2}}%
\left(%
K_n\left(\lambda_2,\lambda_2\right)K_n\left(\lambda,\mu\right)-%
K_n\left(\lambda,\lambda_2\right)K_n\left(\lambda_2,\mu\right)
\right)%
\,d \lambda_2%
\\
+\dfrac{i(n-1)}{2n^2}K_n\left(\lambda,\mu\right)
=-\dfrac{1}{2}V'\left(\lambda\right)
\mathcal{\widetilde{K}}_n\left(x,y\right)
\\
+\dfrac{1}{2n} \intd\limits_{-n\pi}^{n\pi} %
\cot \left(\dfrac{x-z}{2n}\right) %
\left( %
\mathcal{\widetilde{K}}_n\left(z,z\right) \mathcal{\widetilde{K}}_n\left(x,y\right)%
-
\mathcal{\widetilde{K}}_n\left(x,z\right) \mathcal{\widetilde{K}}_n\left(z,y\right)%
\right)%
\, d z
\\
+\dfrac{i(n-1)}{2n}\mathcal{\widetilde{K}}_n\left(x,y\right).\hskip5cm
\end{multline}

\begin{lem}\label{VPLemma}
Denote
\begin{equation*}
D\left(\lambda\right)=V'\left(\lambda\right)+v.p.%
\intd\limits_{-\pi}^{\pi} \cot \dfrac{s}{2}
\rho_n\left(\lambda+s\right) \, ds.
\end{equation*}
Then for any $d>0$ we have uniformly in $\left[a+d,b-d\right]$
\begin{equation*}
\left|D\left(\lambda\right)\right| \leq C n^{-1/4} \ln n.
\end{equation*}
\end{lem}\smallskip

The definition of $\mathcal{K}_n$~\eqref{KerDef}, the above Lemma,
and the bound~\eqref{KerEst} yield
\begin{equation}\label{KerDiff2}
\dfrac{\partial}{\partial x} \mathcal{K}_n\left(x,y\right) = %
\dfrac{1}{2n} v.p.\intd\limits_{-n \pi}^{n \pi} \cot
\left(\dfrac{z-x}{2n}\right) \mathcal{K}_n\left(x,z\right)
\mathcal{K}_n\left(z,y\right) \, dz +O(n^{-1/4} \ln n).
\end{equation}
Below we take $\left|x\right|,\left|y\right|\leq \mathcal{L}=\ln
n$. Then from the inequality $\left|z\right|\leq n\pi$ we get
$\left|\dfrac{x-z}{2n}\right| \leq 3\pi/4$. The function $x\cot x$
is bounded on $\left[0,3\pi/4\right]$, thus
\begin{equation*}
\left|\dfrac{1}{2n}\cot \left(\dfrac{x-z}{2n}\right)\right|\leq C
\left|\dfrac{1}{x-z}\right|.
\end{equation*}
For $\left|x\right|,\left|y\right|\leq \mathcal{L}$ we can
restrict integration in~\eqref{KerDiff2} by the domain
$\left|z\right| \leq 2 \mathcal{L}$, substituting $O(n^{-1/4} \ln
n)$ by $O\left(\mathcal{L}^{-1}\right)$. This follows from the
bound
\begin{gather*}
\left| \dfrac{1}{2n} %
\intd\limits_{2 \mathcal{L} \leq \left|z\right| \leq n \pi} %
\cot \left(\dfrac{x-z}{2n}\right) %
\mathcal{K}_n\left(x,z\right) \mathcal{K}_n\left(z,y\right) \,dz%
\right| %
\\
\leq C \mathcal{L}^{-1} %
\intd \left|\mathcal{K}_n\left(x,z\right)\right|
\left|\mathcal{K}_n\left(z,y\right)\right| \, dz %
\leq C \mathcal{L}^{-1}. %
\end{gather*}
Note that
\begin{equation*}
\dfrac{1}{2n} \cot \dfrac{x}{2n} - \dfrac{1}{x}=%
O\left(n^{-2} \ln n\right), \quad \mbox{for} \, x=O\left(\ln
n\right).
\end{equation*}
Hence,  from the above estimates and~\eqref{KerDiff2} we get
\begin{equation}\label{KerEq}
\dfrac{\partial}{\partial x} \mathcal{K}_n\left(x,y\right) = %
v.p.\intd\limits_{\left|z\right|\leq 2\mathcal{L}}%
\dfrac%
{\mathcal{K}_n\left(x,z\right) \mathcal{K}_n\left(z,y\right)}%
{z-x}%
 \, dz
+O\left(\mathcal{L}^{-1}\right).
\end{equation}
The following lemma shows that $\mathcal{K}_n$ behaves almost like a
difference kernel.\medskip

\begin{lem}\label{KerLemma}
For any $d>0$ we have uniformly in $\lambda_0 \in
\left[a+d,b-d\right]$ and $\left|x\right|,\left|y\right| \leq
nd/4$
\begin{gather}
\left|%
\dfrac{\partial}{\partial x}%
\mathcal{K}_n\left(x,y\right)%
+%
\dfrac{\partial}{\partial y}%
\mathcal{K}_n\left(x,y\right)%
\right|%
\leq %
C%
\left(%
n^{-1/8}+\left|x-y\right| n^{-2}
\right),%
\label{KerLemma1}%
\\
\left|%
\mathcal{K}_n\left(x,y\right)%
-%
\mathcal{K}_n\left(0,y-x\right)%
\right|%
\leq%
C \left|x\right|%
\left(%
n^{-1/8}+\left|x-y\right| n^{-2}
\right).%
\label{KerLemma2}%
\end{gather}
\end{lem}\setcounter{rema}{7}\setcounter{lem}{8}\smallskip

\begin{rema}
Note that the last inequality with $\lambda_0+x_1/n$ instead of
$\lambda_0$, and $x_2-x_1$ instead of $x$ and $y$, leads to the
bound that is valid for any $\left|x_{1,2}\right|\leq nd_0/8$
\begin{equation}\label{LipKer}
\left|%
\mathcal{K}_n\left(x_2,x_2\right)%
-%
\mathcal{K}_n\left(x_1,x_1\right)%
\right|%
\leq C n^{-1/8}%
\left|x_2-x_1\right|.
\end{equation}
\end{rema}\smallskip

\begin{lem}\label{KerH}
For any $\left|x\right|,\left|y\right| \leq \mathcal{L}$
\begin{equation}\label{KerHEq}
\left|%
\dfrac{\partial}{\partial x} %
\mathcal{K}_n\left(x,y\right)%
\right|%
\leq C,%
\quad%
\intd\limits_{\left|x\right|\leq\mathcal{L}}%
\left|%
\dfrac{\partial}{\partial x} %
\mathcal{K}_n\left(x,y\right)%
\right|^2%
\, dx \leq C.
\end{equation}
\end{lem}\smallskip

 Denote
\begin{eqnarray}
\mathcal{K}_{n}^{\ast }(x)
&=&\mathcal{K}_{n}(x,0)\mathbf{1}_{|x|\leq
\mathcal{L}}+\mathcal{K}_{n}(\mathcal{L},0)(1+\mathcal{L}-x)\mathbf{1}_{%
\mathcal{L}<x\leq \mathcal{L}+1}  \label{k^*} \\
&+&\mathcal{K}_{n}(-\mathcal{L},0)(1+\mathcal{L}+x)\mathbf{1}_{-\mathcal{L}%
-1\leq x<-\mathcal{L}},  \notag
\end{eqnarray}%
and observe that for $y=0$ and for any $|x|\leq \mathcal{L%
}/3$, similarly to~\eqref{KerEq}, we can restrict the integration in~\eqref{KerEq} to $%
|z|\leq 2\mathcal{L}/3$ with a mistake $O(%
\mathcal{L}^{-1})$. This and Lemma~\ref{KerLemma} give us the
equation
\begin{equation}
\displaystyle\frac{\partial }{\partial x}\mathcal{K}_{n}^{\ast
}(x)=\int\limits_{|z|\leq 2\mathcal{L}/3}\frac{\mathcal{K}_{n}^{\ast
}(z)\mathcal{K}_{n}^{\ast
}(x-z)}{z}dz+r_{n}(x)+O(\mathcal{L}^{-1}),  \label{K^*Eq}
\end{equation}%
where
\begin{equation*}
r_{n}(x)=\int\limits_{|z|\leq 2\mathcal{L}/3}\frac{\mathcal{K}%
_{n}(z,0)(\mathcal{K}_{n}(x,z)-\mathcal{K}%
_{n}(0,x-z))}{z}dz,
\end{equation*}%
and by Lemma~\ref{KerLemma}, for $|x|\leq \mathcal{L}/3$ we have
\begin{equation*}
r_{n}(x)=O(n^{-1/8}\log n).
\end{equation*}%
Now, using the estimates similar to~\eqref{KerEq}, we can restrict
the integration in~\eqref{K^*Eq} to the real axis. From
Lemma~\ref{KerH} and the relations~\eqref{KerOrt}, \eqref{KerEst}
we get
\begin{equation}
\displaystyle\int |\mathcal{K}_{n}^{\ast }(x)|^{2}dx\leq \displaystyle\int |%
\mathcal{K}_{n}(x,0)|^{2}dx+C^{\prime }\leq C,\quad \displaystyle\int \bigg|%
\displaystyle\frac{d}{dx}\mathcal{K}_{n}^{\ast
}(x)\bigg|^{2}dx\leq C. \label{K^*Est}
\end{equation}%
Consider the Fourier transform
\begin{equation*}
\widehat{\mathcal{K}}_{n}^{\ast }(p)=\int \mathcal{K}_{n}^{\ast
}(x)e^{ipx}dx,
\end{equation*}%
where the integral is defined in the $L^{2}(\mathbb{R})$ sense, and write $%
\mathcal{K}_{n}^{\ast }(x)$ as
\begin{equation}
\mathcal{K}_{n}^{\ast }(x)=(2\pi )^{-1}\int
\widehat{\mathcal{K}}_{n}^{\ast }(p)e^{-ipx}dp.  \label{FInv}
\end{equation}%
From \eqref{StrConv} we have
\begin{equation}
\int \widehat{\mathcal{K}}_{n}^{\ast }(p)dp=2\pi \rho (\lambda
_{0})+o(1), \label{Kprho}
\end{equation}%
and from~\eqref{K^*Est} and the Parseval equation we obtain%
\begin{equation}
\int p^{2}|\widehat{\mathcal{K}}_{n}^{\ast }(p)|^{2}dp\leq C.
\label{pKpEst}
\end{equation}%
From the definition of $\mathcal{K}%
_{n}(x,y)$ we get that the kernel is positive definite
\begin{equation*}
\int\limits_{-\mathcal{L}}^{\mathcal{L}}\mathcal{K}_{n}(x,y)f(x)\overline{f}%
(y)dxdy\geq 0,\quad f\in L_{2}(\mathbb{R}),
\end{equation*}%
therefore from~\eqref{KerLemma2} we have for any function $f\in
L_{2}(\mathbb{R})$
\begin{equation}
\int \widehat{\mathcal{K}}_{n}^{\ast }(p)|\hat{f}(p)|^{2}dp\geq
-C||f||_{L^{2}(\mathbb{R})}^{2}(n^{-1/8}\log
^{4}n+O(\mathcal{L}^{-1})). \label{KpIntEst}
\end{equation}%
From the Parseval equation and~\eqref{KerLemma2} there follows
\begin{equation}
\int |\widehat{\mathcal{K}}_{n}^{\ast
}(p)-\widehat{\mathcal{K}}_{n}^{\ast }(-p)|^{2}dp \leq 2\pi \int
|\mathcal{K}_{n}^{\ast }(x)-\mathcal{K}_{n}^{\ast }(-x)|^{2}dx\leq
Cn^{-1/8}\log ^{3}n.  \label{KpSymm}
\end{equation}%
By the definition of singular integrals
\begin{equation}
\int \frac{\mathcal{K}_{n}^{\ast }(z)\mathcal{K}_{n}^{\ast
}(x-z)}{z}dz=\lim_{\varepsilon \rightarrow
+0}\int dz\mathcal{K}_{n}^{\ast }(z)\mathcal{K}%
_{n}^{\ast }(y-z)\Re (z+i\varepsilon )^{-1}. \label{SingInt}
\end{equation}%
In accordance with the relation
\begin{equation*}
\int e^{ipz}\Re (z+i\varepsilon )^{-1}dz=\pi ie^{-\varepsilon |p|}%
\hbox{sgn}\,p
\end{equation*}%
and the Parseval equation, we can write the r.h.s.
of~\eqref{K^*Eq} as
\begin{multline}\label{RHS_K^*}
\frac{i}{4\pi }\lim_{\varepsilon \rightarrow +0}\int dpdp^{\prime }\widehat{%
\mathcal{K}}_{n}^{\ast }(p)\widehat{\mathcal{K}}_{n}^{\ast
}(p^{\prime })e^{-ipx}\hbox{sign}(p-p^{\prime })e^{-\varepsilon
|p-p^{\prime }|}
 \\
=\frac{i}{2\pi }\int dpe^{-ipx}\widehat{\mathcal{K}}_{n}^{\ast
}(p)\int\limits_{0}^{p}\widehat{\mathcal{K}}_{n}^{\ast }(p^{\prime })dp^{\prime }%
\\ -\frac{i}{4\pi }\int dpe^{-ipx}\widehat{\mathcal{K}}_{n}^{\ast
}(p)\int\limits_{0}^{\infty }(\widehat{\mathcal{K}}_{n}^{\ast }(p^{\prime })-%
\widehat{\mathcal{K}}_{n}^{\ast }(-p^{\prime }))dp^{\prime }.\quad
\end{multline}%
Note that both  integrals are absolutely convergent because $\widehat{%
\mathcal{K}}_{n}^{\ast }\in L^{1}(\mathbb{R})$ by~\eqref{pKpEst}.
Now, using the  Schwarz inequality and~\eqref{pKpEst}, we can
estimate the second component
\begin{multline*}
\left\vert \int\limits_{0}^{\infty }(\widehat{\mathcal{K}}_{n}^{\ast
}(p^{\prime })- \widehat{\mathcal{K}}_{n}^{\ast }(-p^{\prime }))dp^{\prime
}\right\vert \leq \left\vert
\int\limits_{0}^{\mathcal{L}^{2}}(\hat{\mathcal{K}}_{n}^{\ast }(p^{\prime
})-\widehat{\mathcal{K}}_{n}^{\ast }(-p^{\prime }))dp^{\prime }\right\vert
\\
+\int\limits_{|p|>\mathcal{L}^{2}}|\widehat{\mathcal{K}}_{n}^{\ast
}(p^{\prime })|dp^{\prime } \leq \mathcal{L}\left( \int
|\widehat{\mathcal{K}}_{n}^{\ast }(p^{\prime
})-\widehat{\mathcal{K}}_{n}^{\ast }(-p^{\prime })|^{2}dp^{\prime }\right)
^{1/2}+C\mathcal{L}^{-1}.
\end{multline*}%
Thus, from~\eqref{KpSymm}--\eqref{RHS_K^*} we have uniformly in
$|x|<\mathcal{L}/3$
\begin{equation*}
\int \frac{\mathcal{K}_{n}^{\ast }(z)\mathcal{K}_{n}^{\ast
}(x-z)}{z}dz=\frac{i}{2\pi }\int dp\widehat{%
\mathcal{K}}_{n}^{\ast }(p)e^{-ipx}\int\limits_{0}^{p}\widehat{\mathcal{K}}%
_{n}^{\ast }(p^{\prime })dp^{\prime }+O(\mathcal{L}^{-1}).
\end{equation*}%
This allows us to transform~\eqref{K^*Eq} into the following
asymptotic relation that is valid for $|x|\leq \mathcal{L}/3$:
\begin{equation}
\int \widehat{\mathcal{K}}_{n}^{\ast }(p)\bigg(\int\limits_{0}^{p}\widehat{\mathcal{%
K}}_{n}^{\ast }(p^{\prime })dp^{\prime }-p\bigg)e^{-ipx}dp=O(\mathcal{L}%
^{-1}).  \label{KpEq}
\end{equation}%
Consider the functions
\begin{equation}
F_{n}(p)=\int\limits_{0}^{p}\widehat{\mathcal{K}}_{n}^{\ast }(p^{\prime
})dp^{\prime }.  \label{FnKc}
\end{equation}%
Since $p\widehat{\mathcal{K}}_{n}^{\ast }(p)\in L^{2}(\mathbb{R})$,
the sequence $\{F_{n}(p)\}$ consists of functions that are uniformly
bounded and equicontinuous on $\mathbb{R}$. Thus $\{F_{n}(p)\}$ is a
compact family with respect to uniform convergence. Hence, the limit
$F$ of any subsequence $\{F_{n_{k}}\}$ possesses the properties:

\begin{itemize}
\item[(a)] $F$ is bounded and continuous;

\item[(b)] $F(p)=-F(-p)$ (see \eqref{KpSymm});

\item[(c)] $F(p)\leq F(p^{\prime })$, if $p\leq p^{\prime }$ (see \eqref{KpIntEst}%
);

\item[(d)] $F(+\infty )-F(-\infty )=2\pi \rho (\lambda _{0})$ (see \eqref%
{Kprho});

\item[(e)] the following equation is valid for any smooth function
$g$ with the compact support (see \eqref{KpEq}):
\begin{equation}
\int (F(p)-p)g(p)dF(p)=0.  \label{FEq}
\end{equation}
\end{itemize}
The last property implies that $F(p)=p$ or $F(p)=\hbox{const}$, hence it
follows from (a)--(c) that
\begin{equation*}
F(p)=p\,\mathbf{1}_{|p|\leq p_{0}}+p_0\,\hbox{sign}(p)\,\mathbf{1}%
_{|p|\geq p_{0}},
\end{equation*}%
where $p_{0}=\pi \rho (\lambda _{0})$ from (d).
We conclude that \eqref{FEq} is uniquely solvable, thus the sequence $%
\{F_{n}\}$ converges uniformly on any compact to the above $F$. This and (%
\ref{FnKc}) imply the weak convergence of the sequence $\{\mathcal{K}%
_{n}^{\ast }\}$ to the function
$\rho\left(\lambda_0\right)S\left(\rho\left(\lambda_0\right)x\right)$,
where $S(x)$ is defined in \eqref{sin}. But weak convergence
combined with~\eqref{KerEst} and~\eqref{KerHEq} implies
the uniform convergence of $\{\mathcal{K}_{n}^{\ast }\}$ to $\mathcal{K}%
^{\ast }$ on any interval. Thus we have uniformly in $(x,y)$,
varying on a compact set of $\mathbb{R}^2$,
\begin{equation*}
\lim_{n\rightarrow \infty
}\mathcal{K}_{n}(x,y)=\rho\left(\lambda_0\right)S\left(\rho\left(\lambda_0\right)(x-y)\right).
\end{equation*}%
Recalling all definitions, we conclude that Theorem~\ref{MainThm}
is proved.

\begin{center}
\subsection*{Auxiliary Results for Theorem~\ref{MainThm}}
\end{center}

{P r o o f \ of Lemma~\ref{LemmaPolinoms}.} %
Denote
\begin{equation} \label{RDef}
r_{k,j}^{\left(n\right)}=\intd\limits_{-\pi}^{\pi} e^{i \lambda}
\psi_k^{\left(n\right)}\left(\lambda\right)
\overline{\psi_j^{\left(n\right)}\left(\lambda\right)}\, d
\lambda.
\end{equation}
Note that from the orthogonality~\eqref{Ort} we have
$r_{k,j}^{\left(n\right)}=0$ for $j>k+1$. Thus,
\begin{equation}\label{Series}
e^{i \lambda} \, \psi_k^{\left(n\right)}\left(\lambda\right)=%
\sumd\limits_{j=0}^{k+1}r_{k,j}^{\left(n\right)}
\psi_j^{\left(n\right)}\left(\lambda\right).
\end{equation}
Multiplication on $e^{i \lambda}$ is isometric in
$L_2\left[-\pi,\pi\right]$, therefore
\begin{equation*}
\sumd\limits_{j=0}^{k+1}\left| r_{k,j}^{\left(n\right)} \right|^2=%
\left\| \psi_k^{\left(n\right)}\left(\lambda\right) \right\|_2=1.
\end{equation*}
Finally we are ready to prove \eqref{1OneVariable}
\begin{multline}\label{CrDr}
\intd\limits_{-\pi}^{\pi} \left(e^{i \lambda}-e^{i \mu}\right)
\left|K_n\left(\lambda,\mu\right)\right|^2 \, d \mu%
\\
=%
e^{i \lambda}K_n\left(\lambda,\lambda\right)-%
\intd\limits_{-\pi}^{\pi} e^{i \mu}%
\sumd\limits_{m=0}^{n-1}%
\psi_m^{\left(n\right)}\left(\mu\right)%
\overline{\psi_m^{\left(n\right)}\left(\lambda\right)}%
\sumd\limits_{l=0}^{n-1}%
\psi_l^{\left(n\right)}\left(\lambda\right)%
\overline{\psi_l^{\left(n\right)}\left(\mu\right)}%
\, d \mu %
\\
=%
e^{i \lambda}K_n\left(\lambda,\lambda\right)-%
\sumd\limits_{l,m=0}^{n-1}%
r_{m,l}^{\left(n\right)}%
\psi_l^{\left(n\right)}\left(\lambda\right)%
\overline{\psi_m^{\left(n\right)}\left(\lambda\right)}%
\\
=%
r_{n-1,n}^{\left(n\right)}%
\psi_{n-1}^{\left(n\right)}\left(\lambda\right)%
\overline{\psi_n^{\left(n\right)}\left(\lambda\right)}.%
\end{multline}
Now, using the Cauchy inequality and the bound
$\left|r_{n-1,n}^{\left(n\right)}\right| \leq 1$, we get
\eqref{1OneVariable}. Similarly, it is easy to obtain the relation
\begin{equation*}
\intd \limits_{-\pi}^{\pi} %
\left| e^{i \lambda} - e^{i \mu} \right|^2 %
\left| K_n\left( \lambda , \mu \right) \right|^2 %
\, d \mu %
=%
2 \Re
\left\{%
e^{i \lambda}%
r_{n-1,n}^{\left(n\right)}%
\overline{\psi_{n-1}^{\left(n\right)}\left(\lambda\right)}%
\psi_n^{\left(n\right)}\left(\lambda\right)%
\right\} ,
\end{equation*}
which implies \eqref{2OneVariable}. The bounds
\eqref{Variation},\eqref{IntDist1},\eqref{IntDist2} are evident
consequences of~\eqref{2OneVariable}. The lemma is proved.
\hfill\rule{0.5em}{0.5em}\medskip

 %
{P r o o f \ of Lemma~\ref{RhoEstimation}.} Observe that
\begin{equation*}
\dfrac{d\rho_n\left( \lambda \right)}{d\lambda}= \left.
\dfrac{d\rho_n\left( \lambda + t \right)}{d t}\right|_{t=0}.
\end{equation*}
Changing variables in~\eqref{Marginal} $\lambda_j=\mu_j+t$, in
view of periodicity of all functions in the consideration, we have
the representation for $\rho_n\left(\lambda+t\right)$
\begin{equation*}
\rho_n\left(\lambda+t\right) =%
\dfrac{1}{Z_n} \intd e^{-nV\left(\lambda+t\right)}
\prod\limits_{2 \leq j < k \leq n} %
\left| e^{i \mu_j} -  e^{i \mu_k}\right|^2 %
\prod\limits_{j=2}^{n} e^{-nV\left(\mu_j+t\right)} %
\left| e^{i \lambda} -  e^{i \mu_j}\right|^2 %
d \mu_j.
\end{equation*}
After differentiating with respect to $t$, for $t=0$ we get
\begin{multline}\label{DRho}
\dfrac{d\rho_n\left(\lambda\right)}{d\lambda}=
-nV'\left(\lambda\right)p_1^{\left(n\right)}\left(\lambda\right)-
n\left(n-1\right)\intd\limits_{-\pi}^{\pi}
V'\left(\mu\right)p_2^{\left(n\right)}\left(\lambda,\mu\right) d\mu\\
=-V'\left(\lambda\right)K_n\left(\lambda,\lambda\right)-
\intd\limits_{-\pi}^{\pi} V'\left(\mu\right) \left[
K_n\left(\lambda,\lambda\right)K_n\left(\mu,\mu\right)- %
\left| K_n\left(\lambda,\mu\right)\right|^2 \right] d\mu.
\end{multline}
Since $V'\left(\lambda\right)$ is an odd function, and
$K_n\left(\lambda,\lambda\right)$ is an even function, we obtain
\begin{equation*}
\intd\limits_{-\pi}^{\pi}V'\left(\lambda\right)
K_n\left(\lambda,\lambda\right) d\lambda=0.
\end{equation*}
Thus, from~\eqref{DRho} we get
\begin{equation}\label{DRho2}
\rho'_n\left(\lambda\right)=\intd\limits_{-\pi}^{\pi} \left(
V'\left(\mu\right)-V'\left(\lambda\right)
\right) %
\left| K_n\left(\lambda,\mu\right)\right|^2  d\mu.
\end{equation}
We split this integral in two parts corresponding to the domains
$\left| \mu - \lambda \right| \leq d/2$ and $\left| \mu - \lambda
\right| \geq d/2$. In the second integral we
use~\eqref{IntDist1}. It follows from~\eqref{cond1} that in the
first integral we can rewrite $V'\left(\lambda\right)$ as
\begin{multline*}
V'\left(\mu\right)-V'\left(\lambda\right)=\left(\mu - \lambda\right)
V''\left(\lambda\right)+O\left(\left|\mu - \lambda\right|^2\right)
\\
= \left(e^{i \mu} - e^{ i \lambda}\right)
\dfrac{V''\left(\lambda\right)}{i e^{i \lambda}}+O\left(\left(e^{
i \mu} - e^{ i \lambda}\right)^2\right),
\end{multline*}
and using~\eqref{1OneVariable} and~\eqref{2OneVariable}, we
get~\eqref{DRho_bound}. To prove~\eqref{Rho_bound} we use the following
well-known inequality.\smallskip\setcounter{prop}{9}

\begin{prop}
For any function~$u \, : \left[ a_1,b_1\right]\rightarrow
\mathbb{C}$ with $u' \in L_1(a_1,b_1)$ we have
\begin{equation}\label{NIneq}
\left\| u \right\|_{\infty} \leq %
\left\| u' \right\|_1+(b_1-a_1)^{-1}\left\| u \right\|_1,
\end{equation}
where $\| \cdot \|_1, \| \cdot \|_{\infty}$ are the $L_1$ and
uniform norms on the interval $\left[ a_1,b_1\right]$.
\end{prop}\smallskip

This inequality can be obtained easily from the relation
\begin{equation*}
u \left( \lambda \right)=\dfrac{1}{b_1-a_1}
\intd\limits_{a_1}^{b_1}
\left( %
u \left( \lambda \right) -
u \left( \mu \right) %
\right) %
\,d \mu %
+%
\dfrac{1}{b_1-a_1}%
\intd\limits_{a_1}^{b_1}%
u \left( \mu \right) %
\,d \mu. %
\end{equation*}
Using~\eqref{NIneq} for $u=\rho_n$ and the interval $\left[a+d,b-d\right]$,
we get~\eqref{Rho_bound}.\hfill\rule{0.5em}{0.5em}\smallskip

 %
{P r o o f \ of Lemma~\ref{RhoRep}.} From~\eqref{WConv} and~\eqref{sqeqf}
we have for nonreal $z$
\begin{equation}\label{sqeq1}
f^2\left(z\right)-2iV'\left(\mu\right)f\left(z\right)-2iQ\left(z\right)-1=0,
\end{equation}
where $f\left(z\right)$ is the Germglotz transformation of the limiting
density $\rho\left(\lambda\right)$. By~\eqref{Qdefine} and~\eqref{GerInv},
$Q\left(\mu+i0\right)$ is an imaginary valued, bounded, continuous
function. And from~\eqref{GerInv} we obtain
\begin{equation*}
\rho\left(\mu\right)=\dfrac{1}{2\pi}\Re f\left(\mu+i0\right).
\end{equation*}
Computing imaginary and real parts in~\eqref{sqeq1}, we get the
relations
\begin{equation}\label{ImF}
\Im f\left(\mu+i0\right)=V'\left(\mu\right),
\end{equation}
\begin{equation}\label{ReF}
\Re f\left(\mu+i0\right)=\sqrt{2iQ\left(\mu\right)+1-
\left(V'\left(\mu\right)\right)^2},
\end{equation}
from which we obtain~\eqref{SqRep}.\hfill\rule{0.5em}{0.5em}\smallskip

 %
{P r o o f \ of Lemma~\ref{IntRhoEst}.} To prove~\eqref{IntRhoE} with
$k=n-1$ we introduce the probability density
\begin{equation}\label{P_Def}
p_n^{-} \left( \lambda_1, \ldots , \lambda_
{n-1}\right)=\dfrac{1}{Z_n^{-}}\prod\limits_{1 \leq j < k \leq n-1} %
\left| e^{i \lambda_j} -  e^{i \lambda_k}\right|^2 %
\exp \left\{ -n \sum\limits_{j=1}^{n-1} V \left( \lambda_j \right)
\right\}.
\end{equation}
Denote
\begin{equation}\label{Rho_Def}
\rho_n^{-}\left(\lambda\right)=\dfrac{n-1}{n} %
\intd p_n^{-} \left( \lambda,\lambda_2 \ldots ,
\lambda_{n-1}\right) d\lambda_2 \ldots
d\lambda_{n-1}=\dfrac{1}{n}\sumd\limits_{j=0}^{n-2} %
\left| \psi_j^{\left(n\right)}\left(\lambda\right)\right|^2.
\end{equation}
Thus we get
\begin{equation}\label{RhoDiff}
\left| \psi_{n-1}^{\left(n\right)}\left(\lambda\right)\right|^2=%
n \left( \rho_n
\left(\lambda\right)-\rho_n^{-}\left(\lambda\right)\right).
\end{equation}
Analogously to the equation~\eqref{SquareEq}, we can obtain the
"square" equation
\begin{equation}\label{SquareEq_}
\dfrac{i}{2}\left[f_{n}^{-}\left(z\right)\right]^2+%
\intd\limits_{-\pi}^{\pi}%
\dfrac {e^{i \lambda}+e^{i z}}{e^{i \lambda}-e^{i z}}
V'\left(\lambda\right)\rho_n^{-}\left(\lambda\right)
d\lambda=\dfrac{i}{2}+O\left(n^{-2}\eta^{-4}\right),
\end{equation}
for the Germglotz transformation $f_{n}^{-}\left(z\right)$ of the
function~$\rho_n^{-}\left(\lambda\right)$. Denote
\begin{equation}\label{DeltaDef}
\Delta_n\left(z\right)=n\left(
f_n\left(z\right)-f_n^{-}\left(z\right)\right)=%
\intd\limits_{-\pi}^{\pi}%
\dfrac {e^{i \lambda}+e^{i z}}{e^{i \lambda}-e^{i z}}%
\left| \psi_{n-1}^{\left(n\right)}\left(\lambda\right)\right|^2%
d \lambda.
\end{equation}
Subtracting~\eqref{SquareEq_} from \eqref{SquareEq}, we obtain
for~$z=\mu+in^{-1/4}$
\begin{equation*}
\dfrac{i}{2}\Delta_n\left(z\right)\left(
f_n\left(z\right)+f_n^{-}\left(z\right)\right)=
-\intd\limits_{-\pi}^{\pi}\dfrac {e^{i \lambda}+e^{i z}}{e^{i \lambda}-e^{i z}}%
V'\left(\lambda\right)%
\left| \psi_{n-1}^{\left(n\right)}\left(\lambda\right)\right|^2%
d \lambda+O\left(1\right),
\end{equation*}
\begin{multline*}
\dfrac{i}{2}\Delta_n\left(z\right)\left(
f_n\left(z\right)+f_n^{-}\left(z\right)-2iV'\left(\mu\right)\right)
\\
=
\intd\limits_{-\pi}^{\pi}\dfrac {e^{i \lambda}+e^{i z}}{e^{i \lambda}-e^{i z}}%
\left(V'\left(\mu\right)-V'\left(\lambda\right)\right)%
\left| \psi_{n-1}^{\left(n\right)}\left(\lambda\right)\right|^2%
d \lambda+O\left(1\right)=O\left(1\right).
\end{multline*}
Note that $\Re f_n^{-}\left(z\right) > 0$ for $\Im z>0$ therefore
\begin{equation*}
\Re \Delta_n\left(\mu+in^{-1/4}\right)\leq\dfrac{C}{\Re%
f_n\left(\mu+in^{-1/4}\right)}
\end{equation*}
Analogously to~\eqref{fConv}, we can obtain for $z=\mu+in^{-1/4}$
\begin{equation*}
\dfrac{1}{2\pi} \Re
f_n\left(z\right)=\rho\left(\mu\right)+O\left(n^{-1/8}\right)\rho^{-1}\left(\mu\right)
,
\end{equation*}
hence $\Re f_n\left(z\right) \geq C_2$ for sufficiently large $n$, where
$C_2$ is defined in~\eqref{cond1}. Thus,
\begin{equation*}
\Re \Delta_n\left(\mu+in^{-1/4}\right)\leq C.
\end{equation*}
Note that
\begin{equation*}
\Re \dfrac{e^{i \lambda}+e^{i z}}{e^{i \lambda}-e^{i
z}}=\dfrac{\sinh \eta}{\cosh\eta-\cos\left(\mu-\lambda\right)}\geq
C \dfrac{\eta}{\eta^2+\left(\mu-\lambda\right)^2} ,
\end{equation*}
for $\eta^2+\left(\mu-\lambda\right)^2<1$. Thus,
\begin{multline*}
\intd\limits_{\left|\lambda-\mu\right|<n^{-1/4}} %
\left| \psi_{n-1}^{\left(n\right)}\left(\lambda\right)\right|^2 %
\, d \lambda \leq 2 n^{-1/2} \intd\limits_{\left|\lambda-\mu\right|<n^{-1/4}} %
\dfrac{\left| \psi_{n-1}^{\left(n\right)}\left(\lambda\right)\right|^2} %
{n^{-1/2} + \left(\mu-\lambda\right)^2} \, d \lambda
\\
\leq C n^{-1/4} \Re \Delta_n\left(\mu+in^{-1/4}\right)\leq C
n^{-1/4}.
\end{multline*}
A similar bound can be obtained
for~$\psi_{n}^{\left(n\right)}\left(\lambda\right)$ by using the
densities:
\begin{gather*}
p_n^{+}
\left(\lambda_1,\ldots,\lambda_{n+1}\right)=\dfrac{1}{Q_{n,2}^{+}}\prodd\limits_{1
\leq j \leq n+1} e^{-n V(\lambda_j)}\prodd\limits_{1\leq j < k
\leq n+1 } \left| e^{i \lambda_j} - e^{i \lambda_k}\right|^2,
\\
\rho_n^+\left(\lambda\right)=\dfrac{n+1}{n}\intd p_n^{+}
\left(\lambda,\lambda_2,\ldots,\lambda_{n+1}\right) %
\, d\lambda_2 \ldots
d\lambda_{n+1}=\dfrac{1}{n}\sumd\limits_{j=0}^{n} %
\left|\psi_{j}^{\left(n\right)}\left(\lambda\right)\right|^2 \, .
\end{gather*}
Analogously, we will have
$\left|\psi_{n}^{\left(n\right)}\left(\lambda\right)\right|^2=
n\left(\rho_n^+\left(\lambda\right)-\rho_n\left(\lambda\right)\right)$.
Thus, the estimate~\eqref{IntRhoE} is proved. Now we proceed to prove
~\eqref{PolEst} for $k=n$. We use the inequality\medskip

\begin{prop}
For any $C^1$ function~$u \, : \left[ a_1,b_1\right]\rightarrow
\mathbb{C}$
\begin{equation}\label{SobIneq}
\left\| u \right\|^2_{\infty} \leq 2 \left\| u \right\|_2 \left\| u'
\right\|_2+(b_1-a_1)^{-1}\left\| u \right\|^2_2,
\end{equation}
where $\| \cdot \|_2, \| \cdot \|_{\infty}$ are the $L_2$ and
uniform norms on the interval $\left[ a_1,b_1\right]$.
\end{prop}\smallskip

This inequality is a simple consequence of the relation
\begin{equation*}
u^2 \left( \lambda \right)=\dfrac{1}{b_1-a_1}
\intd\limits_{a_1}^{b_1}
\left( %
u^2 \left( \lambda \right) -
u^2 \left( \mu \right) %
\right) %
\,d \mu %
+%
\dfrac{1}{b_1-a_1}%
\intd\limits_{a_1}^{b_1}%
u^2 \left( \mu \right) %
\,d \mu .
\end{equation*}
Consider the interval $\Delta=\left[\lambda - n^{-1/4},\lambda +
n^{-1/4}\right]$ and the function
$\psi\left(\lambda\right)=\psi_{n}^{\left(n\right)}\left(\lambda\right)$.
From the inequality we have
\begin{equation}\label{PsiIneq}
\left|\psi\left(\lambda\right)\right|^2 \leq %
2 \left\| \psi \right\|_{2,\Delta}%
\left\| \psi '  \right\|_{2,\Delta} +%
\dfrac{1}{2} n^{1/4} \left\| \psi\right\|_{2,\Delta} ,
\end{equation}
where $\left\| \cdot \right\|_{2,\Delta}$ is $L_2$~norm on the
interval $\Delta$. It is easy to see that
\begin{equation*}
\left\| \psi \right\|_{2,\Delta} \leq \left\| \psi
\right\|_{2,\left[-\pi,\pi\right]} =1.
\end{equation*}
Denote
$P\left(\lambda\right)=P_n^{\left(n\right)}\left(\lambda\right)$
and $\omega\left(\lambda\right)=e^{-nV\left(\lambda\right)/2}$,
then
$\psi\left(\lambda\right)=P\left(\lambda\right)\omega\left(\lambda\right)$.
Now we estimate $\left\|\psi'\right\|_{2,\left[-\pi,\pi\right]}$:
\begin{gather*}
\left\|\psi'\right\|_{2,\left[-\pi,\pi\right]} =
\left\|P'\omega+P\omega'\right\|_{2,\left[-\pi,\pi\right]} \leq
\left\|P'\omega\right\|_{2,\left[-\pi,\pi\right]}+%
\left\|P\omega'\right\|_{2,\left[-\pi,\pi\right]},
\\
\left\|P\omega'\right\|_{2,\left[-\pi,\pi\right]}=%
\dfrac{n}{2}\left\|PV'\omega\right\|_{2,\left[-\pi,\pi\right]}%
\leq C n \left\|P\omega\right\|_{2,\left[-\pi,\pi\right]}%
= C n,
\end{gather*}
\begin{multline*}
\left\|P'\omega\right\|_{2,\left[-\pi,\pi\right]}^2=\intd
P'\left(\lambda\right) \overline{P'\left(\lambda\right)}
\omega^2\left(\lambda\right) \, d\lambda=%
-\intd P\left(\lambda\right) \overline{P''\left(\lambda\right)}
\omega^2\left(\lambda\right) \, d\lambda  %
\\%
+n \intd P\left(\lambda\right) \overline{P'\left(\lambda\right)}
V'\left(\lambda\right)\omega^2\left(\lambda\right) \, d\lambda.
\end{multline*}
Using the orthogonality
\begin{equation}\label{Ort}
\intd e^{-i m \lambda} \omega \left(\lambda\right) \psi_k^{(n)} \,
d \lambda=0, \quad \mbox{for} \quad m<k,
\end{equation}
we obtain
\begin{gather*}
\intd P\left(\lambda\right) \overline{P''\left(\lambda\right)}
\omega^2\left(\lambda\right) \, d\lambda = %
\intd P\left(\lambda\right) \gamma_n^{(n)} \left(- i n\right)^2 e^{ -i n
\lambda} \omega^2\left(\lambda\right) \, d\lambda\\
= -i n \intd P\left(\lambda\right) \overline{P'\left(\lambda\right)}
\omega^2\left(\lambda\right) \, d\lambda, %
\end{gather*}
where $\gamma_n^{(n)}$ is defined in~\eqref{KInt}. Thus,
\begin{equation*}
\left\|P'\omega\right\|_{2,\left[-\pi,\pi\right]}^2= n \intd
P\left(\lambda\right) \overline{P'\left(\lambda\right)}
\left(V'\left(\lambda\right)+i\right)\omega^2\left(\lambda\right)
\, d\lambda \leq C n
\left\|P'\omega\right\|_{2,\left[-\pi,\pi\right]} ,
\end{equation*}
and we obtain that
$\left\|P'\omega\right\|_{2,\left[-\pi,\pi\right]} \leq C n$.
Combining all above bounds, we conclude that
 $\left\|\psi'\right\|_{2,\left[-\pi,\pi\right]}~\leq~C n$. Now,
using~\eqref{PsiIneq} and~\eqref{IntRhoE}, we obtain~\eqref{PolEst} for
$k=n$. For $k=n-1$ the proof is the same.
\hfill\rule{0.5em}{0.5em}\smallskip

 %
{P r o o f \ of Lemma~\ref{VPLemma}.} Similarly to~\eqref{sqeqf} for
$\eta=n^{-3/8}$ and $\mu \in \left[a+d,b-d\right]$ for $f_n$, defined
in~\eqref{Fdef}, we obtain
\begin{equation}\label{VP1}
\left| %
\Im f_n \left(\mu + i \eta\right)-V'\left(\mu\right) %
\right| \leq C n^{-3/8} \ln n.
\end{equation}
Moreover, we estimate $M=\Im f_n\left(\mu + i\eta\right) +
v.p.\intd\limits_{-\pi}^{\pi} \cot \dfrac{s}{2} \rho_n(\mu +s) \, d
s$. Note that
\begin{equation*}
\Im \dfrac{e^{i \lambda}+e^{i z}}{e^{i \lambda}-e^{iz}}
=-\dfrac{\sin \left(\lambda-\mu\right)}{\cosh \eta- \cos
\left(\lambda-\mu\right)}.
\end{equation*}
Hence,
\begin{multline*}
M=v.p. \intd \left(\cot \dfrac{s}{2} - \dfrac{ \sin s}{\cosh \eta
- \cos s}\right) \rho_n\left(\mu+s\right) \, d s%
\\
=%
\intd\limits_{\left|s\right|\leq d/2} \ln \left( \dfrac{\cosh \eta
- \cos s}{1 - \cos s}\right) \rho_n'\left(\mu+s\right)\, ds +
O\left(\eta\right)=I_1+I_2+I_3+O\left(\eta\right),
\end{multline*}
where $I_1$ is the integral over $\left|s\right|\leq n^{-2}$,
$I_2$ is the integral over $n^{-2}\leq\left|s\right|\leq n^{-1/4}$
and $I_3$ is the integral over $n^{-1/4}\leq\left|s\right|\leq
d/2$. We estimate every term:
\begin{equation*}
\left| I_1 \right|\stackrel{\eqref{PolEst}}{\leq} %
C n^{7/8} \intd\limits_{\left|s\right|\leq n^{-2}} %
\ln \left( \dfrac{\cosh \eta - \cos s}{1 - \cos s}\right)\, ds%
\leq C n^{-9/8} \ln n,
\end{equation*}
\begin{equation*}
\left|I_2\right|\leq C \ln n
\intd\limits_{n^{-2}\leq\left|s\right|\leq n^{1/4}}
\left|\rho_n'\left(\mu+s\right)\right| \, ds
\stackrel{\eqref{IntRhoE}}{\leq} C n^{-1/4} \ln n,
\end{equation*}
\begin{equation*}
\left|I_3\right|\stackrel{\eqref{DRho_bound}}{\leq} C n^{-1/4}
\intd\limits_{\left|s\right|\leq d/2} %
\left(\left| \psi_{n}^{ \left( n \right) } \left( \mu+s \right) \right|^2%
+%
\left| \psi_{n-1}^{ \left( n \right) } \left( \mu+s \right) \right|^2%
\right) \, ds \leq C n^{-1/4}.
\end{equation*}
Combining the above bounds with~\eqref{VP1}, we obtain that the lemma is
proved. \hfill\rule{0.5em}{0.5em}\smallskip

 %
{P r o o f \ of Lemma~\ref{KerLemma}.} To simplify notations we denote for
$t \in \left[0,1\right]$
\begin{equation}\label{LxLyDef}
\lambda_x=\lambda_0+\dfrac{x-tx}{n}, \quad %
\lambda_y=\lambda_0+\dfrac{y-tx}{n}.
\end{equation}
Then, similarly to~\eqref{KerDiff} and~\eqref{DRho2}, we obtain
\begin{equation}\label{DiffKxy}
\dfrac{d}{dt} K_n\left(\lambda_x,\lambda_y\right)= %
x\!\! \intd\limits_{-\pi+\lambda_0}^{\pi+\lambda_0}\!\! %
K_n\left(\lambda_x,\lambda\right)
K_n\left(\lambda,\lambda_y\right)%
\left(%
\dfrac{1}{2}V'\left(\lambda_x\right)+
\dfrac{1}{2}V'\left(\lambda_y\right)- %
V'\left(\lambda\right)
\right)%
\, d\lambda.
\end{equation}
To get our estimates, we split this integral in two parts
$\left|\lambda-\lambda_0\right| \leq d/2$ and
$\left|\lambda-\lambda_0\right| \geq d/2$. By the assumption of
the lemma, $\lambda_x,\lambda_y$ are in
$\left[a+d/2,b-d/2\right]$, thus in the first integral we can
write
\begin{multline*}
V'\left(\lambda\right)-%
\dfrac{1}{2}V'\left(\lambda_x\right)-%
\dfrac{1}{2}V'\left(\lambda_y\right)%
\\
=%
\left(e^{i\lambda}-e^{i\lambda_x}\right)
\dfrac{V''\left(\lambda_x\right)}{2ie^{i \lambda_x}}+
\left(e^{i\lambda}-e^{i\lambda_y}\right)
\dfrac{V''\left(\lambda_y\right)}{2ie^{i \lambda_y}}+
O%
\left(%
\left|e^{i\lambda}-e^{i\lambda_x}\right|^2+%
\left|e^{i\lambda}-e^{i\lambda_y}\right|^2%
\right)%
\\
=%
\left(e^{i\lambda}-e^{i\lambda_x}\right)
\dfrac{V''\left(\lambda_x\right)}{2ie^{i \lambda_x}}+
\left(e^{i\lambda}-e^{i\lambda_y}\right)
\dfrac{V''\left(\lambda_y\right)}{2ie^{i \lambda_y}}
\\
+
O%
\left(%
\left|e^{i\lambda}-e^{i\lambda_x}\right|%
\left|e^{i\lambda}-e^{i\lambda_y}\right|+%
\dfrac{\left|x-y\right|^2}{n^2}%
\right).\hskip3,2cm%
\end{multline*}
Similarly to~\eqref{CrDr}, we obtain
\begin{equation*}
\intd\limits_{-\pi}^{\pi}%
K_n\left(\lambda_x,\lambda\right)
K_n\left(\lambda,\lambda_y\right)%
\left(e^{i \lambda}-e^{i \lambda_x}\right) \, d\lambda=%
-r_{n-1,n}^{(n)} %
\psi_n^{\left(n\right)} \left(\lambda_x\right)%
\overline{\psi_{n-1}^{\left(n\right)} \left(\lambda_y\right)}.
\end{equation*}
Hence,
\begin{equation*}
\intd\limits_{\left|\lambda-\lambda_0\right|\leq d/2}%
\!\! K_n\left(\lambda_x,\lambda\right)
K_n\left(\lambda,\lambda_y\right)%
\left(e^{i \lambda}-e^{i \lambda_x}\right) \, d\lambda = %
-r_{n-1,n} %
\psi_n^{\left(n\right)} \left(\lambda_x\right)%
\overline{\psi_{n-1}^{\left(n\right)} \left(\lambda_y\right)}-I_d,
\end{equation*}
where
\begin{multline*}
\left|I_d\right|=\left|\,\,%
\intd\limits_{\left|\lambda-\lambda_0\right|\geq d/2}%
K_n\left(\lambda_x,\lambda\right)
K_n\left(\lambda,\lambda_y\right)%
\left(e^{i \lambda}-e^{i \lambda_x}\right) \, d\lambda\right| %
\\
\leq%
C \left[\,\,%
\intd\limits_{\left|\lambda-\lambda_0\right|\geq d/2}%
\left|K_n\left(\lambda_x,\lambda\right)\right|^2 \, d\lambda%
\intd\limits_{\left|\lambda-\lambda_0\right|\geq d/2}%
\left|K_n\left(\lambda,\lambda_y\right)\right|^2 \, d\lambda%
\right]^{1/2}%
\\
\stackrel{\eqref{IntDist1}}{\leq C}%
\left[
\left| \psi_{n-1}^{ \left( n \right) } \left( \lambda_x \right) \right|^2%
+
\left| \psi_n^{ \left( n \right) } \left( \lambda_x \right) \right|^2%
+
\left| \psi_{n-1}^{ \left( n \right) } \left( \lambda_y \right) \right|^2%
+
\left| \psi_n^{ \left( n \right) } \left( \lambda_y \right) \right|^2%
\right].\hskip1cm
\end{multline*}
The same bounds are valid for the term with the $e^{i \lambda_y}$
instead of $e^{i\lambda_x}$. To estimate other terms, we use the
Schwarz inequality
\begin{multline*}
\intd%
\limits_{\left|\lambda-\lambda_0\right|\leq d/2}%
\left|%
K_n\left(\lambda_x,\lambda\right)
K_n\left(\lambda,\lambda_y\right)%
\left(e^{i \lambda}-e^{i \lambda_x}\right)%
\left(e^{i \lambda}-e^{i \lambda_y}\right)%
\right|%
\, d\lambda
\\
\leq %
\left[%
\intd\limits_{-\pi}^{\pi}%
\left|
K_n\left(\lambda_x,\lambda\right)%
\left(e^{i \lambda}-e^{i \lambda_x}\right)%
\right|^2%
\, d \lambda%
\intd\limits_{-\pi}^{\pi}%
\left|
K_n\left(\lambda,\lambda_y\right)%
\left(e^{i \lambda}-e^{i \lambda_y}\right)%
\right|^2%
\, d \lambda%
\right]^{1/2}%
\\
\stackrel{\eqref{Variation}}{\leq C}%
\left[%
\left| \psi_{n-1}^{ \left( n \right) } \left( \lambda_x \right) \right|^2%
+
\left| \psi_n^{ \left( n \right) } \left( \lambda_x \right) \right|^2%
+
\left| \psi_{n-1}^{ \left( n \right) } \left( \lambda_y \right) \right|^2%
+
\left| \psi_n^{ \left( n \right) } \left( \lambda_y \right) \right|^2%
\right],
\end{multline*}
\begin{equation*}
\intd%
\limits_{\left|\lambda-\lambda_0\right|\leq d/2}%
\left|%
K_n\left(\lambda_x,\lambda\right)
K_n\left(\lambda,\lambda_y\right)%
\right|%
\, d\lambda \leq %
n
\left(%
\rho_n\left(\lambda_x\right)%
+%
\rho_n\left(\lambda_y\right)
\right)%
\leq Cn.
\end{equation*}
In the second integral we use the boundedness of
$V'\left(\lambda\right)$, the Cauchy inequality
$%
\left|%
K_n\left(\lambda_x,\lambda\right)%
K_n\left(\lambda,\lambda_y\right)%
\right|%
\leq%
\left|%
K_n\left(\lambda_x,\lambda\right)%
\right|^2%
+
\left|%
K_n\left(\lambda,\lambda_y\right)%
\right|^2%
$ and ~\eqref{IntDist1}. Thus,
\begin{equation*}
\left|%
\dfrac{d}{dt} K_n\left(\lambda_x,\lambda_y\right)%
\right|%
\end{equation*}
\begin{equation}\label{DKxyEst}
\leq %
C \left|x\right|%
\left[%
\left| \psi_{n-1}^{ \left( n \right) } \left( \lambda_x \right) \right|^2%
+
\left| \psi_n^{ \left( n \right) } \left( \lambda_x \right) \right|^2%
+
\left| \psi_{n-1}^{ \left( n \right) } \left( \lambda_y \right) \right|^2%
+
\left| \psi_n^{ \left( n \right) } \left( \lambda_y \right) \right|^2%
+
\dfrac{\left|x-y\right|}{n}%
\right].%
\end{equation}
Now, using~\eqref{PolEst}, we obtain
\begin{equation}\label{DKEst}
\left|%
\dfrac{d}{dt} K_n\left(\lambda_x,\lambda_y\right)%
\right|%
\leq %
C \left|x\right|%
\left(%
n^{7/8}+\left|x-y\right| n^{-1}
\right).%
\end{equation}
Finally, observing that
\begin{equation*}
\dfrac{\partial}{\partial x}%
\mathcal{K}_n\left(x,y\right)%
+%
\dfrac{\partial}{\partial y}%
\mathcal{K}_n\left(x,y\right)%
= -\left(xn\right)^{-1} e^{-i (n-1)(x-y)/2n}%
\dfrac{d}{dt}%
\left.%
K_n\left(\lambda_x,\lambda_y\right)%
\right|_{t=0},%
\end{equation*}
\begin{equation*}
\mathcal{K}_n\left(x,y\right)%
-%
\mathcal{K}_n\left(0,y-x\right)%
= %
e^{-i (n-1)(x-y)/2n} \cdot
\dfrac{1}{n}%
\left(%
\left.%
K_n\left(\lambda_x,\lambda_y\right)%
\right|_{t=0}%
-%
\left.%
K_n\left(\lambda_x,\lambda_y\right)%
\right|_{t=1}%
\right),
\end{equation*}
and using~\eqref{DKEst}, we conclude that the lemma is
proved.\hfill\rule{0.5em}{0.5em}\smallskip

 %
{P r o o f \ of Lemma~\ref{KerH}.} First, show that for any
$\left|x\right|\leq nd_0/2$ we have the bound
\begin{equation}\label{KerHM}
\intd\limits_{-1}^{1} %
\dfrac%
{%
\mathcal{K}_n\left(x,x\right) \mathcal{K}_n\left(x+t,x+t\right)-
\left|\mathcal{K}_n\left(x,x+t\right)\right|^2
}%
{t^2} \, dt \leq C.
\end{equation}
Denote
\begin{gather}
\Omega_0=\left[-\pi+\lambda_0,\pi+\lambda_0\right]%
,\quad
\Omega_0^{+}=\Omega_0 / \Omega_0^{-},%
\label{OmDef}
\\
\Omega_0^{-}=\left\{ \lambda \in \Omega_0 : %
\left|%
\sin\dfrac{\lambda-\lambda_0}{2}%
\right|%
\leq \sin\dfrac{1}{2n}
\right\} %
=%
\left[\lambda_0-1/n,\lambda_0+1/n\right]%
,%
\notag
\end{gather}
and consider the quantity
\begin{equation}\label{WDef}
W= %
\left<%
\prodd\limits_{j=2}^{n}%
\left|%
1-\dfrac{\sin^2 1/2n}{\sin^2 \left(\lambda_j-\lambda_0\right)/2}%
\right|%
\right>,
\end{equation}
where the symbol $<\ldots>$ denotes the average with respect to
$p_n\left(\lambda_0,\lambda_2,\ldots,\lambda_n\right)$. We will
estimate $W$ from above. To do this we use the relation
\begin{equation*}
1-\dfrac{\sin^2 \dfrac{1}{2n}}{\sin^2 \dfrac{\mu-\lambda}{2}}=%
\dfrac%
{%
\left(e^{i \left(\lambda+1/n\right)}-e^{i \mu}\right)%
\left(e^{i \left(\lambda-1/n\right)}-e^{i \mu}\right)%
}%
{\left(e^{i \lambda}-e^{i \mu}\right)^2},
\end{equation*}
\eqref{PDef} and the Schwarz inequality. We get that $W^2$ is not
larger than the product of two integrals $I_{+}$ and $I_{-}$,
where
\begin{gather*}
I_{\pm}=Z_n^{-1}\intd\limits_{\Omega_0^{n-1}}%
e^{-n V\left(\lambda_0\right)}%
\prodd\limits_{2 \leq j < k \leq n} %
\left| e^{i \lambda_j} -  e^{i \lambda_k}\right|^2\\ %
\times\exp %
\left\{ %
-n \sum\limits_{j=2}^{n} V \left( \lambda_j \right)
\right\}%
\prodd\limits_{j=2}^{n}%
\left|%
e^{i \left(\lambda_0 \pm 1/n\right)} - e^{i\lambda_j}
\right|^2%
\, d\lambda_j.
\end{gather*}
Moreover, the expression
$n\left(V\left(\lambda_0\right)-V\left(\lambda_0 \pm 1/n
\right)\right)$ is bounded in view of~\eqref{cond1}. Hence, from
\eqref{Marg1} we obtain
\begin{equation}\label{WUpEst}
W\leq C %
\rho_n^{1/2}\left(\lambda_0+1/n\right)%
\rho_n^{1/2}\left(\lambda_0-1/n\right)%
\leq C.
\end{equation}
On the other hand, $W$ can be represented as follows:
\begin{equation}\label{WPhi}
W=%
\left<%
\prodd\limits_{j=2}^{n}%
\left(%
\phi_1\left(\lambda_j\right)%
+
\phi_2\left(\lambda_j\right)%
\right)%
\right>,%
\end{equation}
where
\begin{equation}\label{Phi1Def}
\phi_1\left(\lambda\right)=%
\dfrac%
{%
\left(%
\sin^2 \dfrac{1}{2n}%
-\sin^2 \dfrac{\lambda-\lambda_0}{2}%
\right)^2%
}%
{%
\sin^2 \dfrac{1}{2n}%
\sin^2 \dfrac{\lambda-\lambda_0}{2}%
}%
\mathbf{1}_{\Omega^{-}_0},
\end{equation}
\begin{equation}\label{Phi2Def}
\phi_2\left(\lambda\right)=%
\left(%
1-
\dfrac%
{%
\sin^2 \dfrac{\lambda-\lambda_0}{2}%
}%
{%
\sin^2 \dfrac{1}{2n}%
}%
\right)%
\mathbf{1}_{\Omega^{-}_0}%
+%
\left(%
1-
\dfrac%
{%
\sin^2 \dfrac{1}{2n}%
}%
{%
\sin^2 \dfrac{\lambda-\lambda_0}{2}%
}%
\right)%
\mathbf{1}_{\Omega^{+}_0}.%
\end{equation}
Since $0\leq\phi_2\left(\lambda\right)\leq 1$ and
$\phi_1\left(\lambda\right)\geq 0$, it follows from~\eqref{WPhi}
that $W$ can be estimated bellow as
\begin{equation*}
W\geq \left(n-1\right) %
\intd\limits_{\Omega_0}%
\phi_1\left(\lambda\right)%
\left<%
\delta\left(\lambda_2-\lambda\right)%
\exp\left\{\sumd\limits_{j=3}^{n} \ln \phi_2
\left(\lambda_j\right)\right\}
\right>%
\, d\lambda.%
\end{equation*}
Note that %
$%
\left<\delta\left(\lambda_2-\lambda\right)\right>=%
p_2^{\left(n\right)}\left(\lambda_0,\lambda\right)%
$. Therefore the Jensen inequality implies
\begin{multline*}
W\geq%
\left(n-1\right)
\intd\limits_{\Omega_0^{-}} %
\phi_1\left(\lambda\right)%
p_2^{\left(n\right)}\left(\lambda_0,\lambda\right)\\%
\times \exp%
\left\{%
\left<%
\delta\left(\lambda_2-\lambda\right)%
\sumd\limits_{j=3}^{n} \ln \phi_2\left(\lambda_j\right)%
\right>%
\left[%
p_2^{\left(n\right)}\left(\lambda_0,\lambda\right)%
\right]^{-1}%
\right\}%
\, d\lambda%
\\%
=%
\left(n-1\right)
\intd\limits_{\Omega_0^{-}} %
\phi_1\left(\lambda\right)%
p_2^{\left(n\right)}\left(\lambda_0,\lambda\right)\\%
\times\exp%
\left\{%
\left(n-2\right)
\intd\limits_{\Omega_0}%
\ln\phi_2\left(\lambda'\right)%
p_3^{\left(n\right)}\left(\lambda_0,\lambda,\lambda'\right)%
\, d\lambda'%
\left[%
p_2^{\left(n\right)}\left(\lambda_0,\lambda\right)%
\right]^{-1}%
\right\}%
\, d\lambda,%
\end{multline*}
where $p_k^{\left(n\right)}$ is defined in~\eqref{Marginal}.
Using~\eqref{KernelProperties} for $l=2,3$, we have
\begin{multline}\label{P3}
p_3^{\left(n\right)}\left(\lambda_0,\lambda,\lambda'\right)=%
\dfrac{n}{n-2}\rho_n\left(\lambda'\right)%
p_2^{\left(n\right)}\left(\lambda_0,\lambda\right)%
\\%
+%
\Biggl[\dfrac%
{%
2\Re %
\left(%
K_n\left(\lambda_0,\lambda\right)%
K_n\left(\lambda,\lambda'\right)%
K_n\left(\lambda',\lambda_0\right)%
\right)}{n\left(n-1\right)\left(n-2\right)}%
\\
-%
\dfrac%
{%
K_n\left(\lambda_0,\lambda_0\right)%
\left|%
K_n\left(\lambda,\lambda'\right)%
\right|^2%
+%
K_n\left(\lambda,\lambda\right)%
\left|%
K_n\left(\lambda_0,\lambda'\right)%
\right|^2%
}%
{%
n\left(n-1\right)\left(n-2\right)
}\Biggr].%
\end{multline}
By the Cauchy inequality,
\begin{multline*}
2%
\left|%
K_n\left(\lambda_0,\lambda\right)%
K_n\left(\lambda,\lambda'\right)%
K_n\left(\lambda',\lambda_0\right)%
\right|%
\\
\leq2%
K_n^{1/2}\left(\lambda_0,\lambda_0\right)%
K_n^{1/2}\left(\lambda,\lambda\right)%
\left|%
K_n\left(\lambda,\lambda'\right)%
K_n\left(\lambda',\lambda_0\right)%
\right|%
\\
\leq
K_n\left(\lambda_0,\lambda_0\right)%
\left|%
K_n\left(\lambda,\lambda'\right)%
\right|^2%
+%
K_n\left(\lambda,\lambda\right)%
\left|%
K_n\left(\lambda_0,\lambda'\right)%
\right|^2,%
\end{multline*}
we obtain that the second term in~\eqref{P3} is nonpositive, hence
\begin{equation*}
p_3^{\left(n\right)}\left(\lambda_0,\lambda,\lambda'\right)\leq%
\dfrac{n}{n-2}\rho_n\left(\lambda'\right)%
p_2^{\left(n\right)}\left(\lambda_0,\lambda\right).%
\end{equation*}
Taking into account that $\ln \phi_2\left(\lambda'\right)\leq 0$,
finally we get
\begin{equation}\label{WL}
W\geq\left(n-1\right)%
\intd\limits_{\Omega_0^{-}} \phi_1\left(\lambda\right)%
p_2^{\left(n\right)}\left(\lambda_0,\lambda\right)%
\, d\lambda%
\cdot
\exp%
\left\{%
n\intd\limits_{\Omega_0}%
\rho_n\left(\lambda'\right)%
\ln \phi_2\left(\lambda'\right)%
\, d\lambda'
\right\}.%
\end{equation}
Now we will show that the second multiplier in~\eqref{WL} is
bounded from below
\begin{multline*}
n\intd\limits_{\Omega_0}%
\rho_n\left(\lambda'\right)%
\ln \phi_2\left(\lambda'\right)%
\, d\lambda'%
\\
=\left(%
\intd\limits_{\left|s\right|\leq 1}%
+%
\intd\limits_{1\leq\left|s\right|\leq nd_0/2}%
+%
\intd\limits_{nd_0/2\leq\left|s\right|\leq n\pi}
\right)%
\rho_n\left(\lambda_0+s/n\right)%
\ln \phi_2\left(\lambda_0+s/n\right)%
\, ds %
\\%
\geq%
C%
\left(%
\intd\limits_{\left|s\right|\leq 1}%
\ln%
\left(%
1-%
\dfrac{\sin^2 s/\left(2n\right)}{\sin^2 1/\left(2n\right)}%
\right)%
\, ds%
+%
\intd\limits_{1\leq\left|s\right|\leq nd_0/2}%
\ln%
\left(%
1-%
\dfrac{\sin^2 1/\left(2n\right)}{\sin^2 s/\left(2n\right)}%
\right)%
\, ds%
\right)%
\\
+%
\ln%
\left(%
1-%
\dfrac{\sin^2 1/\left(2n\right)}{\sin^2 d_0/4}%
\right)%
\intd\limits_{\left|s\right|\leq n\pi}%
\rho_n\left(\lambda_0+s/n\right)%
\,ds%
\geq%
C%
\left(%
I_1+I_2
\right)%
+O\left(n^{-1}\right).
\end{multline*}
\begin{equation*}
I_1=\!\intd\limits_{0}^{1}\!%
\ln%
\left(%
\dfrac%
{\cos \left(s/n\right) - \cos \left(1/n\right)}%
{1 - \cos \left(1/n\right)}%
\right)%
 ds%
=%
-n\!\intd\limits_{0}^{1/n}\!%
\dfrac{\sin t}{\sin \left(t+1/n\right)}%
\dfrac{t-1/n}{2 \sin \dfrac{t-1/n}{2}}%
\, dt%
\geq -C
\end{equation*}
\begin{multline*}
I_2=n\intd\limits_{1/n}^{d_0/2}%
\ln %
\left(%
\dfrac%
{\cos \left(1/n\right) - \cos t}%
{1 - \cos t}%
\right)%
\, dt%
=%
\left(nd_0/2-1\right)%
\ln \left(1-\dfrac{\sin^2 1/2n}{\sin^2 d_0/2}\right)
\\
-%
n\left(1-\cos 1/n\right)%
\intd\limits_{1/n}^{d_0/2}%
\cot t/2%
\dfrac{t-1/n}{2 \sin \dfrac{t-1/n}{2}}%
\dfrac{1}{\sin \dfrac{t+1/n}{2}}%
\, dt%
\\%
\geq%
-C-Cn^{-1}\intd\limits_{1/n}^{d_0/2}%
\dfrac{dt}{t\left(t+1/n\right)}%
\geq -C.%
\end{multline*}
Thus, from~\eqref{WUpEst} and~\eqref{WL} we obtain
\begin{equation}\label{P2Est}
n\intd\limits_{\Omega_0^{-}}%
\phi_1\left(\lambda\right)%
p_2^{\left(n\right)} \left(\lambda_0,\lambda\right) \,
d\lambda\geq -C.
\end{equation}
Then,
using~\eqref{KernelProperties},~\eqref{KerDef},~\eqref{Rho_bound},~\eqref{Phi1Def},
and the inequality $\dfrac{1}{t^2} \leq C \dfrac{\sin^2
1/2n}{\sin^2t/2n}$, we obtain~\eqref{KerHM} for $x=0$
from~\eqref{P2Est}. Substituting $\lambda_0$ by $\lambda_0+x/n$,
we get~\eqref{KerHM} for any $\left|x\right|\leq nd_0/2$.

 Now we are ready to prove~\eqref{KerHEq}. Denote
$C_n=\sup\left|\dfrac{\partial}{\partial
x}\mathcal{K}_n\left(x,y\right)\right|$. In view of~\eqref{KerEq}
\begin{align*}
C_n &\leq%
\left|%
\left(%
v.p.%
\intd\limits_{\left|z-x\right|\leq 1}%
+%
\intd\limits_{\left|z-x\right|\geq 1}%
\right)%
\dfrac%
{\mathcal{K}_n\left(x,z\right)\mathcal{K}_n\left(z,y\right)}%
{z-x}%
\, dz%
\right|%
+o\left(1\right)%
\\
&\leq%
\left|I_1\left(x,y\right)\right|
+\left|I_2\left(x,y\right)\right|+o\left(1\right).
\end{align*}
Using the Schwarz inequality and~\eqref{KerOrt}
with~\eqref{KerEst}, we can estimate $I_2$ as follows:
\begin{equation*}
\left|I_2\left(x,y\right)\right| \leq%
\mathcal{K}_n^{1/2}\left(x,x\right)
\mathcal{K}_n^{1/2}\left(y,y\right)\leq C.
\end{equation*}
To estimate $I_1$ denote
\begin{gather}
\hat t_n^*=%
\sup%
\left\{%
t > 0: \left|x-y\right| \leq t \Rightarrow
\mathcal{K}_n\left(x,y\right) \geq \rho_n(\lambda_0)/2 \right\},
\notag
\\
t_n^*=\min \left\{\hat t_n^*,1\right\}. %
\label{T*Def}
\end{gather}
We will prove that the sequence $t_n^*$ is bounded from below by
some nonzero constant. Represent $I_1$ in the form
\begin{align*}
I_1\left(x,y\right)&=%
v.p.%
\intd\limits_{\left|t\right|\leq t_n^*}%
\dfrac%
{%
\mathcal{K}_n\left(x,x+t\right)\mathcal{K}_n\left(x+t,y\right)%
- %
\mathcal{K}_n\left(x,x\right)\mathcal{K}_n\left(x,y\right)%
}%
{t}%
\, dt%
\\%
&+%
\intd\limits_{t_n^*\leq\left|t\right|\leq 1}%
\dfrac%
{%
\mathcal{K}_n\left(x,x+t\right)\mathcal{K}_n\left(x+t,y\right)%
}%
{t}%
\, dt%
=I_1'+I_1''.%
\end{align*}
Using~\eqref{KerEst}, we have $\left|I_1''\right| \leq C \left|\ln
t_n^*\right|$. On the other hand, from~\eqref{Kdef} and the Cauchy
inequality we obtain for any $x,y,z$
\begin{multline}\label{DeltaK}
\left|
\mathcal{K}_n\left(x,z\right)%
-%
\mathcal{K}_n\left(y,z\right)%
\right|^2%
\leq%
\left(%
\mathcal{K}_n\left(x,x\right)%
+%
\mathcal{K}_n\left(y,y\right)%
-%
2%
\mathcal{K}_n\left(x,y\right)%
\right)%
\mathcal{K}_n\left(z,z\right)%
\\
=%
\!\left(%
\!\left(%
\mathcal{K}_n^{1/2}\left(x,x\right)%
- \mathcal{K}_n^{1/2}\left(y,y\right)\!%
\right)^2%
+%
2%
\!\left(%
\mathcal{K}_n^{1/2}\left(x,x\right)%
\mathcal{K}_n^{1/2}\left(y,y\right)%
-%
\mathcal{K}_n\left(x,y\right)\!%
\right)\!%
\right)\!%
\mathcal{K}_n\left(z,z\right).%
\end{multline}
From~\eqref{LipKer} we get that the first term of~\eqref{DeltaK}
is bounded by $C n^{-1/4} \left|x-y\right|^2$. The second term we
rewrite as
\begin{equation*}
\mathcal{K}_n^{1/2}\left(x,x\right)%
\mathcal{K}_n^{1/2}\left(y,y\right)%
-%
\mathcal{K}_n\left(x,y\right)%
=%
\dfrac%
{%
\mathcal{K}_n\left(x,x\right)%
\mathcal{K}_n\left(y,y\right)%
-%
\mathcal{K}_n^2\left(x,y\right)%
}%
{%
\mathcal{K}_n^{1/2}\left(x,x\right)%
\mathcal{K}_n^{1/2}\left(y,y\right)%
+%
\mathcal{K}_n\left(x,y\right)%
}.%
\end{equation*}
Thus, for $\left|x-y\right|\leq t_n^*$ we get
\begin{equation}\label{DeltaK2}
\left|
\mathcal{K}_n\left(x,z\right)%
-%
\mathcal{K}_n\left(y,z\right)%
\right|^2%
\leq%
C%
\left(%
n^{-1/4}\left|x-y\right|^{3/2}%
+%
\mathcal{K}_n\left(x,x\right)%
\mathcal{K}_n\left(y,y\right)%
-%
\left|\mathcal{K}_n\left(x,y\right)\right|^2%
\right).%
\end{equation}
Hence, using~\eqref{DeltaK2},~\eqref{KerHM} and the Schwarz
inequality, we obtain
\begin{gather*}
\left|I_1'\right| \leq C \intd\limits_{\left|t\right|\leq t_n^*}%
\dfrac%
{%
\left|%
\mathcal{K}_n\left(x,x+t\right)%
-%
\mathcal{K}_n\left(x,x\right)%
\right|%
+%
\left|%
\mathcal{K}_n\left(x+t,y\right)%
-%
\mathcal{K}_n\left(x,y\right)%
\right|%
}%
{\left|t\right|} \, dt
\\
\leq%
C\left(t^*_n\right)^{1/2}.%
\end{gather*}
Finally, from the above estimates we have
\begin{equation}\label{CnEst}
C_n \leq C\left(\left|\ln t^*_n\right| +
\left(t^*_n\right)^{1/2}\right).
\end{equation}
Note that if the sequence $t_n^*$ is not bounded from below, then we
have
\begin{equation*}
C\leq\rho_n\left(\lambda_0\right)/2\leq%
\left|\mathcal{K}_n\left(x+t_n^*,x\right)%
-%
\mathcal{K}_n\left(x,x\right)%
\right|%
\leq C_nt_n^* \leq C t_n^* \ln t_n^*+C t_n^*,
\end{equation*}
and we get a contradiction. Thus $t_n^*\geq d^*$ for some
$n$-independent $d^*>0$. Therefore, from~\eqref{CnEst} we obtain
the first inequality of~\eqref{KerHEq}.

 To prove the second inequality of~\eqref{KerHEq}, we
observe that by~\eqref{KerLemma1} we have
\begin{equation*}
\intd\limits_{\left|x\right|\leq\mathcal{L}}%
\left|%
\dfrac{\partial}{\partial x} %
\mathcal{K}_n\left(x,y\right)%
\right|^2%
\, dx=%
\intd\limits_{\left|x\right|\leq\mathcal{L}}%
\left|%
\dfrac{\partial}{\partial y} %
\mathcal{K}_n\left(x,y\right)%
\right|^2%
\, dx+o(1).
\end{equation*}
Then we rewrite the analog of~\eqref{KerEq} for
$\dfrac{\partial}{\partial y} \mathcal{K}_n \left(x,y\right)$ as
\begin{align*}
\dfrac{\partial}{\partial y} \mathcal{K}_n \left(x,y\right)&=%
\left(%
v.p.\intd\limits_{\left|z-y\right|\leq
d*}+\intd\limits_{\left|z\right|\leq
2\mathcal{L}}\mathbf{1}_{\left|z-y\right|\geq d*}
\right)%
\dfrac{\mathcal{K}_n \left(x,z\right)\mathcal{K}_n
\left(z,y\right)}{y-z}\,dz+O\left(\mathcal{L}^{-1}\right) %
\\
&=I_1\left(x,y\right)+I_2\left(x,y\right)+O\left(\mathcal{L}^{-1}\right).
\end{align*}
To complete the proof, it is enough to estimate $I_{1,2}^{2}$.
Since in $I_1$ the domain of integration is symmetric with respect
to $y$, we can write
\begin{align*}
I_1\left(x,y\right)&=%
\intd\limits_{\left|z-y\right|\leq d^*}%
\dfrac{%
\left(%
\mathcal{K}_n \left(x,z\right)-%
\mathcal{K}_n\left(x,y\right)
\right)%
\mathcal{K}_n \left(z,y\right)}{y-z} \,dz%
\\
&+
\intd\limits_{\left|z-y\right|\leq d^*}%
\dfrac{%
\left(%
\mathcal{K}_n \left(z,y\right)-%
\mathcal{K}_n\left(y,y\right)
\right)%
\mathcal{K}_n \left(x,y\right)}{y-z} \,dz.
\end{align*}
Now, using the Schwarz inequality and~\eqref{KerOrt}, we obtain
\begin{align*}
\left|I_1^2\left(x,y\right)\right|&\leq%
2d^*C
\intd\limits_{\left|z-y\right|\leq d^*}%
\dfrac{%
\left|%
\mathcal{K}_n \left(x,z\right)-%
\mathcal{K}_n\left(x,y\right)
\right|^2%
}{\left(z-y\right)^2} \,dz%
\\
&+2d^* \mathcal{K}_n^2 \left(x,y\right)
\intd\limits_{\left|z-y\right|\leq d^*}%
\dfrac{%
\left|%
\mathcal{K}_n \left(z,y\right)-%
\mathcal{K}_n\left(y,y\right)
\right|^2%
}{\left(z-y\right)^2} \,dz.
\end{align*}
Integrating the above inequality with respect to $x$ and
using~\eqref{KerOrt} with~\eqref{KerEst}, we get
\begin{align*}
\intd \left|I_1^2\left(x,y\right)\right| dx%
&\leq C%
\intd\limits_{\left|z-y\right|\leq d^*}%
\dfrac{%
\left|%
\mathcal{K}_n \left(z,y\right)-%
\mathcal{K}_n\left(y,y\right)
\right|^2%
}{\left(z-y\right)^2} \,dz
\\
&+%
C%
\intd\limits_{\left|z-y\right|\leq d^*}%
\dfrac{%
\mathcal{K}_n \left(z,z\right)+%
\mathcal{K}_n\left(y,y\right)-
2\mathcal{K}_n \left(z,y\right)}%
{\left(z-y\right)^2} \,dz.
\end{align*}
Using the bounds~\eqref{DeltaK} in the second integral
and~\eqref{DeltaK2} in the first one, in view of~\eqref{KerHM} we
obtain the bound for $I_1^2$. To estimate $I_2$, we write
\begin{multline*}
\intd \left|I_2^2 \left(x,y\right)\right| dx \leq\!\!\!%
\intd\limits_{\left|z\right|,\left|z'\right|\leq 2\!\!\!\!\!\mathcal{L}}
\mathbf{1}_{\left|z-y\right|>d^*}\mathbf{1}_{\left|z'-y\right|>d^*}
\left|%
\dfrac%
{%
\mathcal{K}_n \left(y,z\right)%
\mathcal{K}_n \left(z,z'\right)%
\mathcal{K}_n \left(z',y\right)%
}%
{%
\left(z-y\right)\left(z'-y\right)
}%
\right|%
dzdz'
\\
\leq C \intd\limits_{\left|z\right|,\left|z'\right|\leq
2\mathcal{L}}
\mathbf{1}_{\left|z-y\right|>d^*}\mathbf{1}_{\left|z'-y\right|>d^*}
\left(
\left|%
\dfrac%
{%
\mathcal{K}_n \left(y,z\right)%
}%
{%
z-y
}%
\right|^2%
+%
\left|%
\dfrac%
{%
\mathcal{K}_n \left(y,z'\right)%
}%
{%
z'-y
}%
\right|^2%
\right)%
dzdz' \leq C.
\end{multline*}
Above bounds for $I_1$ and $I_2$ prove the second inequality
of~\eqref{KerHEq}. Thus, Lemma~\ref{KerH} is proved.
\hfill\rule{0.5em}{0.5em}\smallskip

{\bf Acknowledgement.}
 \ The \ author \ is \ grateful \ to \ Dr. \ M.V. Shcherbina
 \ for \ the problem statement and fruitful discussions.

\label{poplavskyi.tex}


\begin{thebibliography}{99}

\bibitem{Me:91} {\itshape M.L. Mehta},
\newblock {Random Matrices. }
\newblock Acad. Press, New York, 1991.

\bibitem{Kol:97} {\itshape A. Kolyandr},
\newblock {On \ Eigenvalue \ Distribution \ of \ Invariant \ Ensembles \ of \ Random
\ Matrices. ---}
\newblock {\itshape Dop. Ukr. Ac. Sci.} Math. (1997), No.~7, 14--20. (Ukrainian)

\bibitem{Dy:62} {\itshape F.J. Dyson},
\newblock {Statistical Theory of Energy Levels of Complex Systems. I--III. ---}
\newblock {\itshape J.~Math. Phys.} \textbf{3} (1962), 140--175.

\bibitem{Pa-Sh:97} {\itshape L. Pastur and M. Shcherbina},
\newblock {Universality of the Local Eigenvalue Statistics for a~Class of Unitary
Invariant Matrix Ensembles. ---}
\newblock {\itshape J. Stat. Phys.} \textbf{86} (1997), 109--147.

\bibitem{De:99} {\itshape P. Deift},
\newblock {Orthogonal \ Polynomials \ and \ Random \ Matrices: A Riemann--Hilbert Approach ---
CIMS. }
\newblock New York Univ., New York, 1999.

\bibitem{Jo:98} {\itshape K. Johansson},
\newblock {The Longest Increasing Subsequence in a Random Permutation and a Unitary
Random Matrix Model. ---}
\newblock {\itshape Math. Res. Lett.} \textbf{5} (1998), 63--82.

\bibitem{Mu:53} {\itshape N.I. Muskhelishvili},
\newblock {Singular Integral Equations. }
\newblock P. Noordhoff, Groningen, 1953.

\bibitem{Pa-Sh:07} {\itshape L. Pastur and M. Shcherbina},
\newblock {Bulk Universality and Related Properties of Hermitian Matrix Model. ---}
\newblock {\itshape J. Stat. Phys.} \textbf{130} (2007), 205--250.

\bibitem{Dy:72} {\itshape F.J. Dyson},
\newblock {A Class of Matrix Ensembles. ---}
\newblock {\itshape J. Math. Phys.} \textbf{13} (1972), 90--107.

\bibitem{De-Co:99} {\itshape P. Deift, T. Kriecherbauer, K.T.-K.
    McLaughlin, S. Venakides, and X. Zhou},
\newblock {Uniform Asymptotics for Polynomials Orthogonal
    with Respect to Varying Exponential Weights and Applications to
    Universality Questions in Random Matrix Theory. ---}
\newblock {\itshape Comm. Pure Appl. Math.} \textbf{52} (1999), 1335--1425.

\end{thebibliography}
\end{document}